\documentstyle[11pt,epsf]{article}

\def\beq{\begin{eqnarray}}
\def\eeq{\end{eqnarray}}
\def\bea{\begin{eqnarray*}}
\def\eea{\end{eqnarray*}}

\def\NPB#1#2#3{Nucl. Phys. {\bf B#1} (#2) #3}
\def\PLB#1#2#3{Phys. Lett. B {\bf #1} (#2) #3}

\def\PRD#1#2#3{Phys. Rev. {\bf D#1} (#2) #3}

\def\PRL#1#2#3{Phys. Rev. Lett. {\bf #1} (#2) #3}
\def\PREP#1#2#3{Phys. Rep. {\bf #1} (#2) #3}

\def\centeron#1#2{{\setbox0=\hbox{#1}\setbox1=\hbox{#2}\ifdim
\wd1>\wd0\kern.5\wd1\kern-.5\wd0\fi
\copy0\kern-.5\wd0\kern-.5\wd1\copy1\ifdim\wd0>\wd1
\kern.5\wd0\kern-.5\wd1\fi}}
\def\ltap{\;\centeron{\raise.35ex\hbox{$<$}}{\lower.65ex\hbox{$\sim$}}\;}
\def\gtap{\;\centeron{\raise.35ex\hbox{$>$}}{\lower.65ex\hbox{$\sim$}}\;}
\def\gsim{\mathrel{\gtap}}
\def\lsim{\mathrel{\ltap}}

\def\slashchar#1{\setbox0=\hbox{$#1$}           
   \dimen0=\wd0                                 
   \setbox1=\hbox{/} \dimen1=\wd1               
   \ifdim\dimen0>\dimen1                        
      \rlap{\hbox to \dimen0{\hfil/\hfil}}      
      #1                                        
   \else                                        
      \rlap{\hbox to \dimen1{\hfil$#1$\hfil}}   
      /                                         
   \fi}                                        %

\def\singleandabitspaced{\baselineskip=\normalbaselineskip\multiply
    \baselineskip by 120\divide\baselineskip by 100}
\def\singlespaced{\baselineskip=\normalbaselineskip}

\newcommand{\newc}{\newcommand}
\newc{ \Ni         }{ {\tilde N}_i }
\newc{ \Nj         }{ {\tilde N}_j }
\newc{ \NI         }{ {\tilde N}_1 }
\newc{ \NII        }{ {\tilde N}_2 }
\newc{ \NIII       }{ {\tilde N}_3 }
\newc{ \NIIII      }{ {\tilde N}_4 }
\newc{ \Ci         }{ {\tilde C}_i }
\newc{ \Cj         }{ {\tilde C}_j }
\newc{ \CI         }{ {\tilde C}_1 }
\newc{ \CII        }{ {\tilde C}_2 }
\newc{ \CIp        }{ {\tilde C}^{+}_1 }
\newc{ \CIm        }{ {\tilde C}^{-}_1 }
\newc{ \Cip        }{ {\tilde C}^{+}_i }
\newc{ \Cim        }{ {\tilde C}^{-}_i }
\newc{ \Cjp        }{ {\tilde C}^{+}_j }
\newc{ \Cjm        }{ {\tilde C}^{-}_j }
\newc{ \G          }{ {\tilde G} }
\newc{ \XI         }{ {\tilde X}_1 }
\newc{ \XII        }{ {\tilde X}_2 }
\newc{ \eL         }{ {\tilde e}_L }
\newc{ \eR         }{ {\tilde e}_R }
\newc{ \veL        }{ {\tilde \nu} }
\newc{ \SG         }{ {\tilde \gamma} }
\newc{ \SZ         }{ {\tilde Z} }
\newc{ \gmu        }{ \gamma^{\mu} }
\newc{ \gnu        }{ \gamma^{\nu} }
\newc{ \gamone      }{ \gamma_{1} }
\newc{ \gamtwo      }{ \gamma_{2} }
\newc{ \gamboth      }{ \gamma_{1,2} }
\newc{ \dL         }{ \tilde d_L }
\newc{ \dR         }{ \tilde d_R }
\newc{ \uL         }{ \tilde u_L }
\newc{ \uR         }{ \tilde u_R }
\newc{ \slepton    }{ \widetilde \ell }
\newc{ \M          }{ {\cal M} }
\newc{ \ra         }{ \rightarrow }
\newc{ \ltilde     }{ {\tilde \ell} }
\newc{ \nutilde    }{ {\tilde \nu} }
\newc{ \lLstar     }{ { \tilde \ell}_L^* }
\newc{ \lRstar     }{ { \tilde \ell}_R^* }
\newc{ \snu        }{ { \tilde \nu} }
\newc{ \snustar    }{ { \tilde \nu}^* }
\newc{ \nubar      }{ \overline{ \nu } }
\newc{ \muL        }{ { \tilde \mu}_L }
\newc{ \muR        }{ { \tilde \mu}_R }
\newc{ \tauL       }{ { \tilde \tau}_L }
\newc{ \tauR       }{ { \tilde \tau}_R }
\newc{ \h          }{ { h^0 } }
\newc{ \Et         }{ { \slashchar{E}_T } }
\newc{ \Etot         }{ { \slashchar{E} } }
\newc{ \spt         }{ { \slashchar{p}_T } }
\newc{ \Etcut      }{ { \slashchar{E}_T^{\rm cut} } }
\newc{ \sigbreff   }{ \sigma \times {\rm BR} \times {\rm EFF} }
\newc{ \eegg       }{ {ee\gamma\gamma} }
\newc{ \GeV        }{ {\rm GeV} }

\newcommand{\sss}{\scriptscriptstyle}

\newcommand{\stoop}{\tilde{t}_{1}}
\newcommand{\sel}{\tilde{e}_{\sss L}}
\newcommand{\ser}{\tilde{e}_{\sss R}}
\newcommand{\lR}{\tilde{\ell}_{\sss R}}
\newcommand{\lL}{\tilde{\ell}_{\sss L}}
\newcommand{\smur}{\tilde{\mu}_{\sss R}}
\newcommand{\staur}{\tilde{\tau}_{\sss R}}

\newcommand{\gino}{\tilde{g}}
\newcommand{\squark}{\tilde{q}}
\newcommand{\sneutrino}{\tilde{\nu}}

\newcommand{\minvisible}{M_{\rm Invis}}

\begin{document}

\begin{titlepage}
\begin{flushright}
{\large
 hep-ph/9605398 \\
 May 1996 \\
}
\end{flushright}

\vskip 1.2cm

\begin{center}
{\LARGE\bf Search for supersymmetry with a light gravitino}

{\LARGE\bf at the Fermilab Tevatron and CERN LEP colliders}

\vskip 2cm

{\large
 S. Ambrosanio$^{*,}$\footnote{{\tt ambros@umich.edu}},
 G. L. Kane\footnote{{\tt gkane@umich.edu}},
 Graham D. Kribs\footnote{{\tt kribs@umich.edu}},
 Stephen P. Martin\footnote{{\tt spmartin@umich.edu}},
} \\
\vskip 4pt
{\it Randall Physics Laboratory, University of Michigan,\\
     Ann Arbor, MI 48109--1120 } \\
\vskip 10pt
{\large
 S. Mrenna\footnote{{\tt mrenna@hep.anl.gov}}
} \\
\vskip 4pt
{\it High Energy Physics Division, Argonne National Laboratory, \\
     Argonne, IL 60439 } \\

\vskip 1.5cm

\begin{abstract}

We analyze the prospects for discovering supersymmetry at the Fermilab
Tevatron and CERN LEP colliders in the scenario that the lightest
supersymmetric particle is a gravitino of mass $\lsim$
1 keV. We consider in particular the case that the lightest
neutralino has a nearly 100\% branching fraction into gravitino + photon
within the detector. This implies that supersymmetric events
should contain both missing (transverse) energy and two
energetic photons. Therefore one can search for supersymmetry
simply through inclusive production of superpartners.
We consider the exclusion and reach capabilities of the Tevatron in
exploring the supersymmetric parameter space, and study the efficiencies
which can be achieved in this search. We also consider
the discovery reach and backgrounds at LEP with $\sqrt{s} =$ 160, 175,
and 190 GeV.

\end{abstract}

\end{center}

\vskip 1.0 cm

\noindent
$^*$Supported mainly by an INFN postdoctoral fellowship, Italy.

\end{titlepage}
\setcounter{footnote}{0}
\setcounter{page}{2}
\setcounter{section}{0}
\setcounter{subsection}{0}
\setcounter{subsubsection}{0}

\singleandabitspaced
\section*{1. Introduction}
\indent

One of the intriguing theoretical aspects of supersymmetry is that
if it is realized as a local symmetry, it necessarily and automatically
incorporates gravity.
This connection is of no consequence in most studies of
supersymmetric phenomenology at colliders, because of the
familiar negligibility of gravitational interactions. However,
this need not be so if the gravitino (the spin $3/2$ partner of
the graviton) is very light. The gravitino ($\G$) obtains its mass
by absorbing the spin $1/2$ would-be goldstino associated with the
spontaneous breaking of supersymmetry. In the high-energy limit,
the interactions
of the $\pm 1/2$ helicity components of the gravitino are the same
as those of the goldstino it has absorbed. As emphasized originally
by Fayet~\cite{Fayet}, these interactions are proportional to $1/m_\G$
in the $m_\G \rightarrow 0$ limit and are
therefore potentially important even for processes at ordinary energies.

However, the strength of gravitino interactions (or equivalently
$1/m_\G$) certainly
cannot be arbitrarily large. The gravitino mass is related to the scale
of spontaneous supersymmetry breaking by
\beq
m_{\G} = {\Lambda_{\rm SUSY}^2 \over \sqrt{3} M }\> = \>
5.9\times 10^{-5}
\left ({ \Lambda_{\rm SUSY} \over 500 \> {\rm GeV} } \right )^2 \>{\rm eV}\> .
\label{gravmass}
\eeq
Here $M = (8 \pi G_{\rm Newton} )^{-1/2} = 2.4 \times 10^{18}$ GeV,
and $\Lambda_{\rm SUSY}^{-2}$ is the coupling of the would-be goldstino
to the divergence of the supercurrent.
Now, the scale $\Lambda_{\rm SUSY}$ should at least exceed the mass of
the heaviest of the superpartners of the Standard Model (SM) particles,
typically a gluino ($\gino$) or squark ($\squark$),
and probably greatly so.
[One might expect a significant hierarchy between $\Lambda_{\rm SUSY}$
and the electroweak scale, in order that negative radiative corrections
to the Higgs scalar (mass)$^2$ can be effective in driving electroweak
symmetry breaking.] Thus if one takes e.g.~a
bound $\Lambda_{\rm SUSY} > 500 $ GeV,
equation (\ref{gravmass}) becomes a lower
bound on the gravitino mass of roughly $6\times 10^{-5}$ eV.
In any case a given mass spectrum for the sparticles always implies
a lower bound on $m_{\G}$.
This is equivalent to a bound on the strength of
the gravitino's interactions with the SM particles and their
superpartners. This type of bound is quite conservative,
and is certainly not expected to be saturated in particular models
\cite{oldmodels,GaugeMediated} of
supersymmetry-breaking at low energies.
For example, recently proposed models \cite{GaugeMediated}
of dynamical supersymmetry breaking communicated
to the visible sector by gauge interactions evidently favor
$m_\G \gsim 1$ eV
which automatically avoids a dangerous R-axion \cite{Raxion}, although
other ways of doing this might be possible.

One can also attempt to obtain a
lower bound on the gravitino mass by examining the requirements
of partial-wave unitarity for, e.g., the scattering of two gluons
into two gravitinos \cite{BhattacharyaRoy}. That process has
contributions at tree-level
from t- and u-channel exchanges of the gluino. As shown
in ref.~\cite{BhattacharyaRoy}, tree-level partial wave unitarity is
violated in this process when $\sqrt{s}$ exceeds
\beq
E_{\rm crit} = \sqrt{288 \pi } M m_{\G} / m_{\gino} \> .
\label{unitarity-eq}
\eeq
Now, one way to interpret this result
is that the gravitino interactions, being gravitational, should not
become strong below, say, the scale $M$;
then (\ref{unitarity-eq}) would become a lower bound on the
gravitino mass of $m_\G \gsim m_{\gino}/30$. If this were true, the
gravitino would always interact far too weakly to play any role in
collider experiments. However, it seems preferable to interpret the
critical energy indicated by (\ref{unitarity-eq})
as the maximum value of a
scale $\Lambda_{\gino}$ of unknown new physics at which $m_{\gino}$
becomes effective.
In that case, one finds only that
$m_{\G} \gsim \Lambda_{\gino} m_{\gino}/30 M$.
If $\Lambda_{\gino}$ is smaller than the
ultimate scale of supersymmetry breaking $\Lambda_{\rm SUSY}$, this
constraint is vacuous when compared with (\ref{gravmass}).

On the other hand, cosmological constraints \cite{cosmoconstraints}
seem to place an upper
bound on $m_{\G}$ of about $10^4$ eV, at least in the absence of
late inflation. There
is then still a window of perhaps 9 orders of magnitude
for the mass of a light gravitino. In particular classes of models, this window
can be much smaller. Throughout this window, $m_{\G}$
is clearly insignificant for collider kinematics,
and so can be taken to
simply parameterize the strength of the gravitino's interactions.

Most collider phenomenology studies
performed up to now assume that the lightest supersymmetric
particle (LSP) is a neutralino (mixture of neutral higgsinos and
$SU(2)_L \times U(1)_Y$ gauginos). If, as is most
often assumed, $R$-parity is exactly conserved, then supersymmetric
particles will always be produced in pairs, and the LSP is absolutely
stable. In this ``neutralino LSP scenario",
every supersymmetric event will feature two LSPs
leaving the detector. Therefore the signals for supersymmetry always
involve missing energy, often together with lepton and/or multi-jet
signatures corresponding to
particular decay chains of the superpartners produced \cite{review}.
In the ``gravitino
LSP scenario", however, the signatures should be quite different if
the decays of superpartners into gravitinos occur within the detector
a significant fraction of the time.

For example, in the most obvious case that the next-to-lightest
supersymmetric particle (NLSP) is a neutralino ($\NI$) with a
non-zero photino component, one has the interesting
decay \cite{Fayet,decay}
\beq
\NI \rightarrow \gamma \G\> .
\label{n1decay-eq}
\eeq
In the rest frame of the decaying $\NI$, the photon takes an energy
equal to $m_{\NI} /2$ and is produced isotropically.
The photons produced in
supersymmetric events should therefore often be energetic enough to
pass cuts designed to reduce SM backgrounds.
The gravitino still carries away a significant amount
of missing energy. Thus, supersymmetric signals in the gravitino LSP
scenario should be similar to those in the neutralino LSP scenario
but with two (one) energetic, often isolated, photons if both (one)
of the $\NI$ decays
occur within the detector. Several recent papers have presented
interesting studies of this type of signal at the Next Linear Collider
\cite{SWY}, $e\gamma$ colliders \cite{KNW}, and the
Tevatron \cite{DTW}.

Since the presence of
additional energetic photons in supersymmetric events would provide
a welcome and powerful discriminant against SM
backgrounds, it is possible to consider supersymmetry
discovery signals
based on {\em inclusive} production of all superpartners.
The signatures in the gravitino LSP case are
$\gamma\gamma \Et +X$ (or possibly $\gamma \Et +X$),
where $X$ is an arbitrary collection
of leptons+jets (including $X=$ nothing which can occur
e.g.~in the cases of $\NI\NI$ or $\sneutrino\sneutrino$
production). The purpose of the present paper is to study
this strategy at the Fermilab Tevatron and CERN LEP colliders
in the Minimal Supersymmetric Standard Model (MSSM)
with a light gravitino. We define the MSSM to be the
minimal supersymmetric extension of the SM
with $R$-parity conserved.
Squarks (other than the top squarks) are assumed to be
very nearly degenerate in mass, as suggested both by theory and
the absence of flavor-changing neutral currents in experiment.
Sleptons with the same electroweak quantum numbers are also
assumed to be degenerate in mass. We will also often, but not always,
make use of the ``gaugino mass unification"
assumption for running gaugino mass parameters:
\beq
M_2 = {3 \over 5 \tan^2 \theta_W} M_1 = {\alpha_2 \over \alpha_3} M_3
\label{gauginomassunification}
\eeq
which arises both in gauge-mediated and gravity-mediated supersymmetry
breaking.\footnote{In particular, this ``unification"
relation can be well-motivated
even in the absence of gauge-coupling unification at a very high energy scale;
see e.g.~\cite{GaugeMediated}.}

This study is motivated in part by the observation at CDF of a single
$\eegg  \Et$ event \cite{Event} that does
not seem to have a SM interpretation.
It has already been suggested
that this event can be explained by
supersymmetry, either in the gravitino LSP scenario \cite{DDRT,AKKMM}
considered here
or in a scenario with a higgsino-like neutralino LSP \cite{AKKMM}.
In \cite{AKKMM}, we found that if this event is due to selectron
pair production followed by the decays $\tilde e \rightarrow e \NI$ and
$\NI \rightarrow \gamma \G$, then the
kinematic requirements of the event
place rough bounds of 80 GeV $< m_{\tilde e}$ and
38 GeV $< m_{\NI} <$ 100 GeV. However, this event can also be
ascribed to pair production of charginos, as we shall remark below.
In any case, we will maintain a more general point of view in most of
the present work,
rather than restrict our attention to the parameter space suggested by
that one event.

This paper is organized as follows. In section 2, we will consider
the decays of supersymmetric particles into 2- and 3-body final states
which
include a gravitino. The 2-body decays $\NI \rightarrow Z \G$ and
$\NI \rightarrow h^0\G$ (which might compete with $\NI \rightarrow
\gamma\G$ if the photino component of $\NI$ is very small)
turn out to be subject to
a very strong 
kinematic suppression. We also discuss
expectations
for the decay lengths of $\NI$, and note the existence of
regions of parameter
space where the decay $\NI \rightarrow \gamma\G$ is unduly suppressed.
In section 3, we consider the limits on the reach of the Tevatron
by studying the cross-sections for inclusive sparticle production.
These rates are more interesting
than in the case of the neutralino LSP scenario because of the relative
ease with which arbitrary types of sparticle production can be
detected using energetic photons.
We propose a set of cuts designed to maximize the efficiency
for detection of supersymmetry at the Tevatron via the signature
$\gamma\gamma \Et +X$,
and study the efficiencies and lepton and jet multiplicities
obtained using several sets of model parameters as test cases.
In section 4 we discuss the possibility of discovering
supersymmetry at LEP with $\sqrt{s} = 160$, 175, and 190 GeV,
including an analysis of the relevant backgrounds.
Section 5 contains some concluding remarks.

\section*{2. Decays into gravitinos}
\indent

The decay $\NI \rightarrow \gamma\G$ will play the central role
in the phenomenological discussions to follow. In sections 3 and 4 we will
simply assume that this decay is the only important one involving the
gravitino, and that it always occurs within the detector.
However,  it is interesting to consider other possible decays which
might have an impact on collider phenomenology as well.
If $\NI$ is the NLSP and is fairly heavy,
(but especially if it is beyond the reach of the Tevatron
with the present integrated luminosity), the decays $\NI \rightarrow h^0 \G$
and $\NI\rightarrow Z\G$ can at least naively be important.
Also it is interesting to consider the possibility that a sparticle other than
$\NI$ is the NLSP. Finally, if $m_{\G}$ is very small ($\ll 1$ eV)
one can even entertain the possibility that superpartners other
than the NLSP can decay directly into final states containing a
gravitino. In this section we present general formulas for the decay
widths of supersymmetric particles into final states involving gravitinos.

The relevant interactions of the gravitino are given by
\cite{Fayet,sugra}
\beq
{\cal L} \supset {1\over 8M} {\overline \lambda}^A \gamma^\rho \sigma^{\mu\nu}
\G_\rho F_{\mu\nu}^A +
{1\over \sqrt{2} M }\overline \psi_L \gamma^\mu \gamma^\nu \G_\mu
D_\nu \phi + h.c.
\label{lagrangian-eq}
\eeq
where the spin 3/2 gravitino field is $\G_\mu$ and $\lambda^A$
is the gaugino associated with the gauge field contained in the
field strength $F_{\mu\nu}^A$,
and $(\phi,\psi)$ are the scalar and fermionic components of the
chiral supermultiplets. The full gravitino field can be well-approximated
by its spin 1/2 goldstino component when it appears as an external
state in processes at energy scales relevant for collider studies:
\beq
\G_\mu \approx \sqrt{2\over 3} {i\over m_{\G}} \partial_\mu \G\> .
\eeq
In this limit it is not difficult to use eq.~(\ref{lagrangian-eq})
to calculate decay rates of supersymmetric particles in the MSSM
to final states including gravitinos. Let us first consider the decays of
neutralinos. Using the relation
between the mass eigenstates and the gauge eigenstates of the $\Ni$,
one finds the decay widths of $\Ni$ into neutral gauge bosons
to be as follows:
\beq
& &\Gamma (\Ni \rightarrow \gamma \G ) =
{\kappa_{i\gamma} \over 48 \pi}{m_{\Ni}^5 \over M^2 m_{\G}^2}
\label{ngammaGdecay}
\\
& &\Gamma (\Ni \rightarrow Z\G ) =
{2\kappa_{iZ_T}+ \kappa_{iZ_L}\over 96 \pi}
{m_{\Ni}^5  \over M^2 m_{\G}^2}
\left ( 1- {m_Z^2\over m_{\Ni}^2}\right )^4
\label{nzGdecay-eq}
\eeq
where \beq
& &\kappa_{i\gamma} = |N_{i1}\cos\theta_W + N_{i2}\sin \theta_W |^2
\\
& &\kappa_{iZ_T} = |N_{i1}\sin\theta_W - N_{i2} \cos\theta_W |^2
\\
& &\kappa_{iZ_L} = |N_{i3} \cos \beta -  N_{i4}\sin\beta|^2
\eeq
measure the contents in $\Ni$ of photino, zino, and the higgsino
partner of the neutral would-be Nambu-Goldstone boson, respectively.
(Here and in the following we use the notations of \cite{HaberKane} for
the parameters
and mixing matrices of neutralinos and Higgs scalar bosons in the
MSSM. Thus $N_{ij}$ are the neutralino
mixing matrices with $(i,j)$ the (mass, gauge) eigenstate labels, and
$\tan\beta$
is the ratio of Higgs vacuum expectation values.)
In these and similar formulas below, the Planck-scale suppression
$m_{\Ni}^2/M^2$
is numerically counteracted by the hierarchy $m^2_{\Ni}/m^2_{\G}$,
so that the decay width can be non-negligible. So for example we can write
(\ref{ngammaGdecay}) in the suggestive form
\beq
\Gamma(\Ni\rightarrow\gamma\G ) = 1.12\times 10^{-11} \>{\rm GeV}\>
\kappa_{i\gamma}
\left ( m_{\Ni}\over 100\>{\rm GeV} \right )^{5}
\left ( m_{\G} \over 1\> {\rm eV} \right )^{-2}
\> .
\eeq

When the 2-body decay $\Ni \rightarrow Z\G$ is near threshold,
the formula (\ref{nzGdecay-eq}) is not reliable; also when the 2-body
decay is not kinematically allowed, the decay can still proceed through an
off-shell $Z$ boson. In these situations one must use the 3-body
decay formula.
In the limit of massless SM fermions from the off-shell
$Z$, the width of the neutralino from 3-body decays through
a virtual $Z$ boson are obtained by replacing
\beq
(2\kappa_{iZ_T} + \kappa_{iZ_L})
\left ( 1- {m_Z^2\over m_{\Ni}^2}\right )^4 \longrightarrow
2\kappa_{iZ_T} I_1 + \kappa_{iZ_L} I_0
\eeq
in (\ref{nzGdecay-eq}), where the kinematic factors are most compactly
written as
\beq
I_n=
{\epsilon\over \pi}
\int_0^1 dx \> {(1-x)^4\> (x/R)^n \over (x-R)^2 + \epsilon^2 }
\label{funkyintegrals}
\eeq
with $R=m_Z^2/m_{\Ni}^2$ and $\epsilon = \Gamma_Z m_Z/m_{\Ni}^2$.
In the case that $m_{\Ni}-m_Z\gg \Gamma_Z$,
one finds $I_0 \approx I_1 \approx
\left ( 1- {m_Z^2/m_{\Ni}^2}\right )^4$ so that (\ref{nzGdecay-eq})
is recovered. At threshold ($m_{\Ni} \approx m_Z$) one finds
$I_0 \approx 4 I_1 \approx .0029$, rather than 0 as suggested
by the 2-body formula
(\ref{nzGdecay-eq}). For another point of reference,
with $m_{\NI} = 70$ GeV one finds the
kinematic factors are approximately $I_0 \approx .0013$ and
$I_1 \approx .00015$.
In order for off-shell or near-threshold decays
$\NI \rightarrow Z^{(*)} \G$ to compete with
$\NI \rightarrow \gamma \G$, the photino component of $\NI$ clearly would have
to be very small. Even for $m_{\NI}=150$ GeV one finds
$I_0 \approx I_1 \approx 0.16$.

Next we consider the decays of a neutralino into a gravitino plus
any of the neutral Higgs scalar boson mass eigenstates
$\varphi= (h^0,H^0,A^0)$ of the MSSM. The 2-body decay widths are
given by
\beq
\Gamma (\Ni \rightarrow \varphi\G ) =
{\kappa_{i\varphi} \over 96 \pi}{m_{\Ni}^5 \over M^2 m_{\G}^2}
\left ( 1- {m_\varphi^2\over m_{\Ni}^2}\right )^4
\label{nhGdecay-eq}
\eeq
where the relevant higgsino contents are given by
\beq
& &\kappa_{ih^0} = |N_{i3}\sin\alpha -N_{i4}\cos\alpha |^2
\\
& &\kappa_{iH^0} = |N_{i3}\cos\alpha +N_{i4}\sin\alpha |^2
\\
& &\kappa_{iA^0} = |N_{i3}\sin\beta +N_{i4}\cos\beta |^2
\eeq
Note that with the identifications $\varphi = G^0$
(the electroweak would-be Nambu-Goldstone boson),
$\kappa_{iG^0} = |N_{i3}\cos\beta -N_{i4}\sin\beta |^2$
and $m_{G^0} = m_Z$, one recovers the decay width into longitudinal
$Z$'s indicated in equation (\ref{nzGdecay-eq}), in compliance with
the equivalence theorem \cite{Equivalence}.

It is certainly
not an outlandish possibility that $m_{\NI} > m_{h^0}$, so that
the two body decay $\NI \rightarrow h^0\G$ can compete with $\NI
\rightarrow \gamma \G$. However, this decay is also crippled by
the 
kinematic suppression indicated in (\ref{nhGdecay-eq})
unless $m_{\NI}$ is significantly larger than $m_{h^0}$.
The 3-body decay widths for $\Ni \rightarrow \varphi^* \G$
with $\varphi^*$ decaying into pairs of SM fermions
(treated as massless for kinematic purposes) are given by
replacing in (\ref{nhGdecay-eq})
\beq
\left ( 1- {m_\varphi^2\over m_{\Ni}^2}\right )^4 \longrightarrow I_1
\eeq
as given by (\ref{funkyintegrals})
with $R=m_\varphi^2/m_{\Ni}^2$ and $\epsilon = \Gamma_\varphi
m_\varphi/m_{\Ni}^2$. Since in the MSSM the width of $h^0$ is only a few
MeV, such three-body and near-threshold decays are generally negligible.

The decay widths of sleptons and heavy squarks are also easily found.
For a sfermion decaying into a massless SM fermion + gravitino,
one finds the 2-body width
\beq
\Gamma({\tilde f} \rightarrow f \G) =
{m_{\tilde f}^5\over 48 \pi M^2 \tilde m_\G^2}
\eeq
One of the more intriguing possibilities is that the nearly
degenerate right-handed
sleptons $\ser$, $\smur$, $\staur$ act effectively as co-NLSPs.
In that case, the supersymmetry discovery signatures would
generally
involve at least two
energetic leptons + $\Et$. If a sneutrino is the NLSP, then signatures
should often be similar to those
in the neutralino LSP scenario, since the
decays $\tilde \nu \rightarrow \nu \G$ are invisible.
It does not appear to be sensible to contemplate a left-handed
charged slepton
as the NLSP, because of the sum rule
\beq
m^2_{\lL} = m^2_{\tilde \nu} + m_W^2 |\cos 2\beta|
\label{sumrule}
\eeq
for $\tan\beta>1$. Squark (mass)$^2$ parameters receive large positive
contributions proportional to $\alpha^2_s$ and/or $\alpha_s$, so that it
seems doubtful that a squark could be the NLSP. One possible exception
is the lightest top squark mass eigenstate ($\stoop$) since
$m_{\stoop}^2$ can receive large negative radiative corrections
proportional to the top Yukawa coupling squared. However, if
$m_{\stoop} \lsim m_t$,
then $\stoop$ should be very long-lived if it is the
NLSP, and in particular should always hadronize and
escape the detector as a
charge 0 or 1 ``mesino" ($\stoop \overline q$)
or as a charge 0, 1, or 2
``sbaryon" ($\stoop q q^\prime$) bound state. In any case, for the
remainder of this paper, we will
decline to consider the possibility that a sfermion could be the NLSP.

The 2-body decay widths of charginos ($\Ci$)
into gravitino final states are given by
formulas entirely analogous to (\ref{nzGdecay-eq}) and (\ref{nhGdecay-eq}):
\beq
& &\Gamma (\Cip \rightarrow W^+\G ) =
{2\kappa_{iW_T}+ \kappa_{iW_L}\over 96 \pi}
{m_{\Ci}^5  \over M^2 m_{\G}^2}
\left ( 1- {m_W^2\over m_{\Ci}^2}\right )^4
\\
& &\Gamma (\Cip \rightarrow H^+\G ) =
{\kappa_{iH^+}\over 96 \pi}
{m_{\Ci}^5  \over M^2 m_{\G}^2}
\left ( 1- {m_{H^+}^2\over m_{\Ci}^2}\right )^4
\eeq
with
\beq
& &\kappa_{iW_T} = {1\over 2}\left ( | V_{i1} |^2 + |U_{i1} |^2 \right )
\\
& &\kappa_{iW_L} = |V_{i2}|^2 \sin^2 \beta + |U_{i2}|^2 \cos^2\beta
\\
& &\kappa_{iH^+} = |V_{i2}|^2 \cos^2 \beta + |U_{i2}|^2 \sin^2\beta
\eeq
The generalizations to off-shell decays are given by the obvious analogs
of the above expressions for $\Ni$ decays. However, it should be noted
that because of the form of the chargino and neutralino mass matrices,
a chargino can only be the NLSP in a small and not particularly attractive
region of parameter space.

In general, if $m_{\G}$ could be arbitrarily small
compared to superpartner masses, then all decays of supersymmetric
particles could proceed directly to the corresponding SM
particle plus gravitino. However, as a practical matter for
supersymmetric states accessible at Tevatron energies and
taking into account a conservative lower bound on the gravitino
mass as mentioned in the Introduction or stricter bounds in
particular classes of models, it is easy to see that
the decay widths for non-NLSP sparticles
listed above should be quite small and should be overwhelmed by the
usual well-studied
decays. Besides the decay of the NLSP, there is one other
potential exception which seems worthy of mention.
If the gravitino mass is near the lower end of the allowed window,
it is possible that a heavy gluino can decay directly to gluon+gravitino
through a 2-body decay, rather than following the usual cascade
decay pattern through virtual squarks. Since the {\em only} other decays
of the gluino are mediated by
virtual squarks which can be quite heavy in models,
it is conceivable that the direct decay to gravitino can dominate
in the gravitino LSP scenario.
(In contrast, decays of all other non-NLSP superpartners can proceed
through virtual $W$s or $Z$s or sparticles which are plausibly much
lighter than squarks.) The relevant decay width is given by
\beq
\Gamma (\gino \rightarrow g \G) =
{m_{\gino}^5 \over 48 \pi M^2 m_{\G}^2 }
= 1.1 \times 10^{-9} \> {\rm GeV}\>
 \left ({m_{\gino}\over  250\>{\rm GeV}}\right )^5
\left ({m_{\G}\over 1\> {\rm eV}}\right )^{-2}
\label{ggGdecay}
\eeq
The competition between this decay and the usual cascade decays of gluinos
has already been studied in \cite{DicusPRD41,DreesWoodside}, where
(\ref{ggGdecay})
was found to be negligible unless $m_{\G} \lsim 10^{-2}$ eV, even
if all squarks are as heavy as several TeV.
For gravitinos lighter than $10^{-2}$ eV,
it was found that the direct decays (\ref{ggGdecay}) can dominate over
the more conventional decay chains through virtual squarks only if
\beq
m_{\G} \lsim 10^{-3} \>{\rm eV}\> \left (m_{{\tilde q}}\over
1\> {\rm TeV}\right )^2 \> .
\eeq
This can only occur in the slightly problematic case that
$\Lambda_{\rm SUSY}$ does not greatly exceed $m_{\squark}$ [cf.
equation (\ref{gravmass})].

Given the considerations above, we will optimistically assume for the remainder
of this paper that a neutralino is the NLSP, that the branching
fraction for $\NI \rightarrow \gamma\G$ is 100\%, and that all
supersymmetric decay chains terminate in this subdecay. Nevertheless,
it is worth noting that this decay can be strongly suppressed due to
a very small photino content of $\NI$ in regions of parameter space
with small $|\mu|$, making $\NI$ long-lived on collider scales.
Assuming the usual gaugino mass unification condition
(\ref{gauginomassunification}) and restricting our attention to the
parameter space not already excluded by LEP with $\tan\beta > 1.5$,
$\kappa_{1\gamma}$ can be less than $0.001$
only if $\mu$ is negative and $|\mu|/M_2 < 0.2$.
For $\kappa_{1\gamma} < 0.01$, it is required that
$|\mu|/M_2 < $ 0.4 (0.2) for $\mu$ negative (positive).
A milder but still quite significant
suppression $\kappa_{1\gamma} < 0.1$ can be obtained if
$|\mu|/M_2 < $ 0.5 (0.65) for $\mu$ negative (positive).
Conversely, as long as $|\mu| > M_2$ and $m_{\CI}>50$ GeV, one finds
$\kappa_{1\gamma} > 0.21$ (0.13) for $\mu$ negative (positive).
Formally, if $\kappa_{1\gamma}$ were to vanish, the decay
$\NI \rightarrow \gamma\G$ could
still proceed through one-loop graphs, but
these amplitudes are very small \cite{DEJS} and in the
present context are only competitive for
a fine-tuning of $\kappa_{1\gamma}$ which is finer than we are
willing to contemplate here.

Of course, a 100\% branching fraction for $\NI\rightarrow \gamma\G$
does not guarantee that the photons can be detected, since the
length scale associated with this decay might easily be comparable to
the relevant physical size of the detector.
The probability that each $\NI$ with energy
$E$ in the lab frame will travel a distance $\leq x$ before decaying
is given by
\beq
P(x) = 1-e^{-x/L}
\eeq
where from (\ref{ngammaGdecay}) the decay length is
\beq
L = 1.76 \times 10^{-3} \>  (\kappa_{1\gamma})^{-1} \>
({E^2/ m_{\NI}^2} - 1)^{1/2}                          \>
\left ( m_{\NI}\over 100\>{\rm GeV} \right )^{-5}
\left ( m_{\G} \over 1\> {\rm eV} \right )^2 \>{\rm cm}\>.
\label{decleng}
\eeq
Clearly $L$ depends strongly on $m_{\NI}$ and $m_{\G}$ and
can either be larger than, or negligible compared to,
the relevant physical dimension ($\sim 150$ cm) of a CDF-type
detector. Note that if
$L$ is larger than 150 cm, the efficiency for detecting
one photon can greatly exceed that for detecting both. For example, taking
$L$ to be 15 meters, one finds that the probability for both (one)
of the photons being emitted within 150 cm of the event vertex is roughly
0.01 (0.17). For $L\approx 150$ cm, the probability of two (one)
photons being emitted within 150 cm of the event vertex is 0.40 (0.47).
Since the SM backgrounds for events with one
energetic photon greatly exceed those for events with two such photons,
we will optimistically assume in the following discussion that
for the processes of interest $L< 150$ cm, so that all
supersymmetric events will lead to two potentially detectable photons.
Taking $\kappa_{1\gamma}$ and
$(E^2/m_{\NI}^2 -1)^{1/2}$ to be of order unity, this requires roughly $m_{\G}
\lsim 250$ eV for a 100 GeV $\NI$. Larger decay lengths will decrease the
efficiency of detection accordingly.

\section*{3. Supersymmetry with a light gravitino at the Tevatron}
\indent

The presence of two energetic photons from supersymmetric events
in the gravitino LSP scenario should
dramatically increase the detectability over that found in
the usual neutralino LSP scenario.
In this section we
will study the possibility for detecting inclusive $\gamma\gamma\Et+X$
signals at the Tevatron in the present data sample of about 100 pb$^{-1}$ per
detector.  We concentrate on signals from chargino and neutralino,
slepton, and light stop squark
production for a range of models,
and we comment on other potential signals.
For this study, we assume that
the decays
$\NI\rightarrow \gamma\G$ occur within the detector 100\% of the time.
As a practical matter, we compute kinematics of events with the further
assumption that these decays occur close to the event vertex.
All event simulation is performed using the {\tt PYTHIA} Monte Carlo
with supersymmetric extensions \cite{spythia}. Refs.~\cite{SWY,KNW,DTW}
also contain recent studies of gravitino LSP physics at colliders.

\subsection*{A. Chargino and Neutralino Production}
\indent

The production
cross sections for $\Ci\Cj$, $\Ci\Nj$ and $\Ni\Nj$ at hadron
colliders are functions of the gaugino--higgsino parameters
(the $U(1)$ gaugino mass $M_1$, the $SU(2)$ gaugino mass $M_2$,
the supersymmetric Higgsino mass parameter
$\mu$, $\tan\beta$) and the squark masses.
In the following we vary these parameters
to find the range of expected signals.
The gravitino LSP scenario has striking phenomenological implications
for ${\Ci}$ and ${\Ni}$ production.
Consider, for example, the process
$p{\overline p} \rightarrow {\tilde{C}}_{1}^{\pm} \NII$.
In the neutralino LSP
scenario, it is well--known that this process can be detected with a
good efficiency when the final state includes
three leptons \cite{baer}.
However, in the gravitino LSP scenario, all of the
final states of this process (including
$\gamma\gamma \ell^+ \ell^- \ell^{\prime +} \Et$,
$\gamma\gamma  \ell^+ jj\Et$,
$\gamma\gamma \ell^+ \ell^- jj\Et$ and
$\gamma\gamma jjjj\Et$)
can provide useful signals. Likewise, the production of $\CIp\CIm$ pairs
can lead to observable signals
$\gamma\gamma \ell^+ \ell^{\prime -} \Et$,
$\gamma\gamma \ell^\pm jj\Et$, and
$\gamma\gamma jjjj\Et$.
When the gaugino unification condition (\ref{gauginomassunification})
is satisfied,  these two processes provide the
bulk of the supersymmetric signal throughout much of parameter space,
because of the relatively large couplings $W\CI\NII$ and $Z\CI\CI$,
$\gamma\CI\CI$.
Also $\NI\NI$ production,
which is undetectable at hadron colliders
in the neutralino LSP scenario,
leads to the signal $\gamma\gamma\Et$ in the gravitino LSP scenario.
Unfortunately, although this process is kinematically favored, it
usually has a negligible cross section at hadron colliders because of
a small $Z\NI\NI$ coupling and heavy squarks.

\begin{figure}[t]
\centering
\epsfxsize=4in
\hspace*{0in}
\epsffile{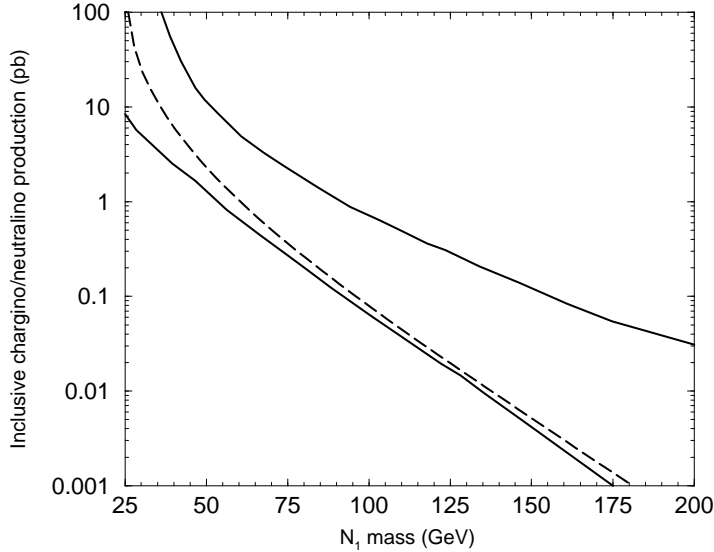}
\caption{Total production cross sections
for charginos and neutralinos
($\protect{\Ni\Nj}$ and $\protect{\Ni\Cj}$ and $\protect{\Ci\Cj}$)
at the Tevatron with
$\protect\sqrt{s} = 1.8 \; \protect{\rm TeV}$ as a function of the
lightest neutralino mass, assuming gaugino
mass unification. The solid lines are the minimum and maximum allowed
cross sections. The dashed line is a typical large $|\mu|$ and
heavy squark
limit ($\protect\mu=1000$ GeV, $\protect{m_{\squark} = 1000}$ GeV,
$\protect{\tan\beta = 1.5}$).}
\label{inofigure1}
\end{figure}
The branching fractions for the various final states associated with
chargino and neutralino production are quite model--dependent.
For example, the jet or lepton multiplicity and kinematics can
be a strong function of squark and slepton masses.
Since all of these final states from ${\Ci}$ and ${\Ni}$ production
involve two energetic photons and missing transverse energy,
we prefer to focus on the inclusive $\gamma\gamma \Et+X$ signal rather
than details of jet or lepton multiplicity.  If a number of events
are found in the data sample which are not understood as coming from
the SM, then such details could help disentangle the
underlying theory.  Below we show such distributions for a
specific set of models.

\begin{figure}[t]
\centering
\epsfxsize=4in
\hspace*{0in}
\epsffile{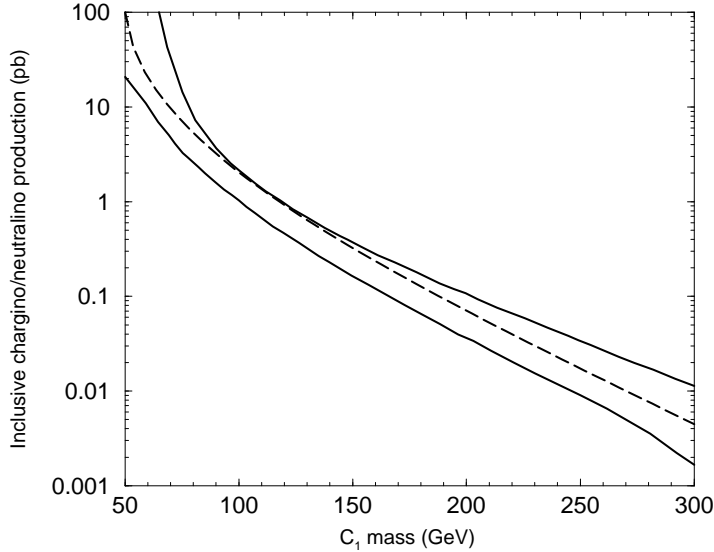}
\caption{As in Fig.~\ref{inofigure1}, but as a function of $m_{\CI}$.
We find that even when the gaugino
mass unification assumption (\ref{gauginomassunification}) is
not made, the minimum cross section is never significantly less than
shown here.}
\label{inofigure2}
\end{figure}

In Fig.~\ref{inofigure1} we show the allowed range for the
total inclusive production cross-section
$\sigma (p{\overline p} \rightarrow
\Ni\Nj$  or $ \Ci\Cj$ or $\Ni \Cj )$
at $\sqrt{s} =$ 1.8 TeV
as a function of the lightest neutralino mass $m_{\NI}$.
We have assumed that the
gaugino mass unification assumption (\ref{gauginomassunification})
holds, so that the gaugino--higgsino sector is determined by only
three parameters, one of which we choose to be the ${\NI}$ mass.
To generate this graph, we have varied
the other parameters of the MSSM over the
ranges
\begin{eqnarray*}
&250 \>{\rm GeV}\> < \>m_{\squark}\> < \> 1000\>{\rm GeV}&\\ \nonumber
&-1000 \>{\rm GeV}\> < \>\mu \> < \> 1000\>{\rm GeV}&\\ \nonumber
&1.5 \> < \>\tan\beta \> < \> 55 &\nonumber
\end{eqnarray*}
The dashed line represents a typical large $|\mu|$, heavy squark limit,
namely $\mu=1000$ GeV, $m_{\squark} = 1000$ GeV, $\tan\beta=1.5$.
If the gaugino mass unification condition (\ref{gauginomassunification})
is not satisfied, then the total inclusive chargino-neutralino
production cross--section (for the range of $m_{\NI}$ shown)
can be essentially negligible; this is
traceable directly to the kinematic suppression associated with
very heavy charginos. In Fig.~\ref{inofigure2}
we show the same cross-section, this
time as a function of the lighter chargino mass.
Again, Fig.~\ref{inofigure2}
assumes (\ref{gauginomassunification}). However,
we found that the minimum production cross-section in the case of
general gaugino mass parameters
is {\em not} significantly lower than that shown
in Fig.~\ref{inofigure2}, for a given $\CI$ mass.
This is important because it shows that the Tevatron can set model-independent
exclusion limits on $m_{\CI}$, if the efficiency for
detection is reasonably bounded from below.
The maximum production cross-section for general gaugino mass
parameters not obeying (\ref{gauginomassunification}) can be several times
larger than that shown in Fig.~\ref{inofigure2}.
As can be seen from Fig.~\ref{inofigure2},
the total number of chargino/neutralino
pair production events at the Tevatron with 100 pb$^{-1}$ of data
is $>100$ before cuts if a chargino mass is less than $100$ GeV
(the maximum LEP reach).

To fully define a signal, we choose the following cuts for
the two photons $\gamma_1,\gamma_2$:

$\bullet$ $E_T(\gamma_1), \> E_T(\gamma_2) > 12$ or $25$ GeV,
where $E_T$ is the transverse energy.
We define two different signals based on the minimal $E_T$.
For the lower threshold, we impose a 30\% loss of efficiency
when one or both photons
have $E_T$ below 25 GeV, to simulate the approximate loss of triggering
efficiency~\cite{bob_blair}.

$\bullet$ $|\eta(\gamma_1)|,\> |\eta(\gamma_2)| < 1$,
where $\eta$ is the pseudorapidity.

$\bullet$ $E_T^{\rm iso}<$ 4 GeV
\footnote{This number can be significantly reduced without affecting
the signal.}, where
\[ E_T^{\rm iso} = \sum_{j,R<0.4}^{}E_T^{(j)}-E_T(\gamma), \]
and we sum the transverse energy from all particles $(j)$ within a cone of size
$R=\sqrt{(\Delta\eta)^2+(\Delta\phi)^2}$ centered on the photon
candidate.  (Photons from jets or bremsstrahlung tend not to be isolated
from additional hadronic activity.)

$\bullet$ $\Et > 30$ GeV, where $\Et$ is determined by the sum of the
visible energy in smeared jets, photons, and leptons.

Standard Model physics
backgrounds can arise from $W^\pm(\to
\ell\nu_\ell)\gamma\gamma$,
$Z(\to \nu{\bar \nu},\tau^{+}\tau^{-})\gamma\gamma$,
and $Q{\bar Q}\gamma\gamma$ where $Q=c,b,$ or $t$.
We have not made a full matrix element simulation of these
backgrounds.  However, we estimate their magnitude by generating
$W^\pm\gamma$, $Z\gamma$, $t\bar t$, $b\gamma X$, and $c\gamma X$ events
with additional QED radiation in the leading--logarithm
approximation \cite{torbjorn}.  Based on this analysis,
we estimate
$(0.13,0.11,<10^{-3},<10^{-2},<0.1)$
events in 100~pb$^{-1}$ from each of these
sources when $E_T(\gamma )>$ 12 GeV.  For this set of cuts, the
$c\gamma\gamma$ background is the largest hadronic source, and it is
well under control.
Likewise, backgrounds from a jet faking an isolated photon can be
estimated from these numbers using a simple scaling by
$\alpha_s/\alpha_{em}\times R_{j\to\gamma}\times f\approx .1$,
where $R_{j\to\gamma}\approx 10^{-3}$ is the probability a jet
fakes a photon and $f\approx 10$ accounts for the squared quark charge.
The probability of two jets faking two photons is even further
suppressed.
Finally, backgrounds with a fake missing transverse energy are limited by
excellent electromagnetic calorimetry.  Essentially, these chosen
cuts
should yield a signal free from background, though the $\Et$ cut
could be increased if necessary.

To assess the sensitivity of the Tevatron to $\gamma\gamma\Et +X$ signals
from ${\Ci}$ and ${\Ni}$ production, we have performed event level
simulations of various light gravitino models using the parameters:
\begin{eqnarray}
  \begin{array}{c}
  M_2 = 100,150,200,225 \>{\rm GeV};                        \\
  \mu = \pm125,\pm215,\pm300,\pm600 \>{\rm GeV};            \\
  \tan\beta = 1.5,\>1.7,\>2.5,\>3.0,\>10.0;                 \\
  (m_{\tilde q},m_{\lL},m_{\lR}) = (250,125,119) {\rm~or~}
  (500,250,238) \>{\rm GeV}.
  \end{array}
\label{testmodels-eq}
\end{eqnarray}
Here $M_1$ is fixed by (\ref{gauginomassunification}).
The sneutrino mass is fixed by the sum rule (\ref{sumrule}).
(When the result is less than the ${\NI}$ mass,
we take instead $m_{\tilde \nu} = m_{{\NI}}+5$ GeV and
$m_{\lL}$ fixed by the sum rule.)
While larger squark masses can easily be obtained in models,
we find that to a good approximation, the dependence of signals
on squark mass
\begin{figure}
\centering
\epsfxsize=3.4in
\hspace*{0in}
\epsffile{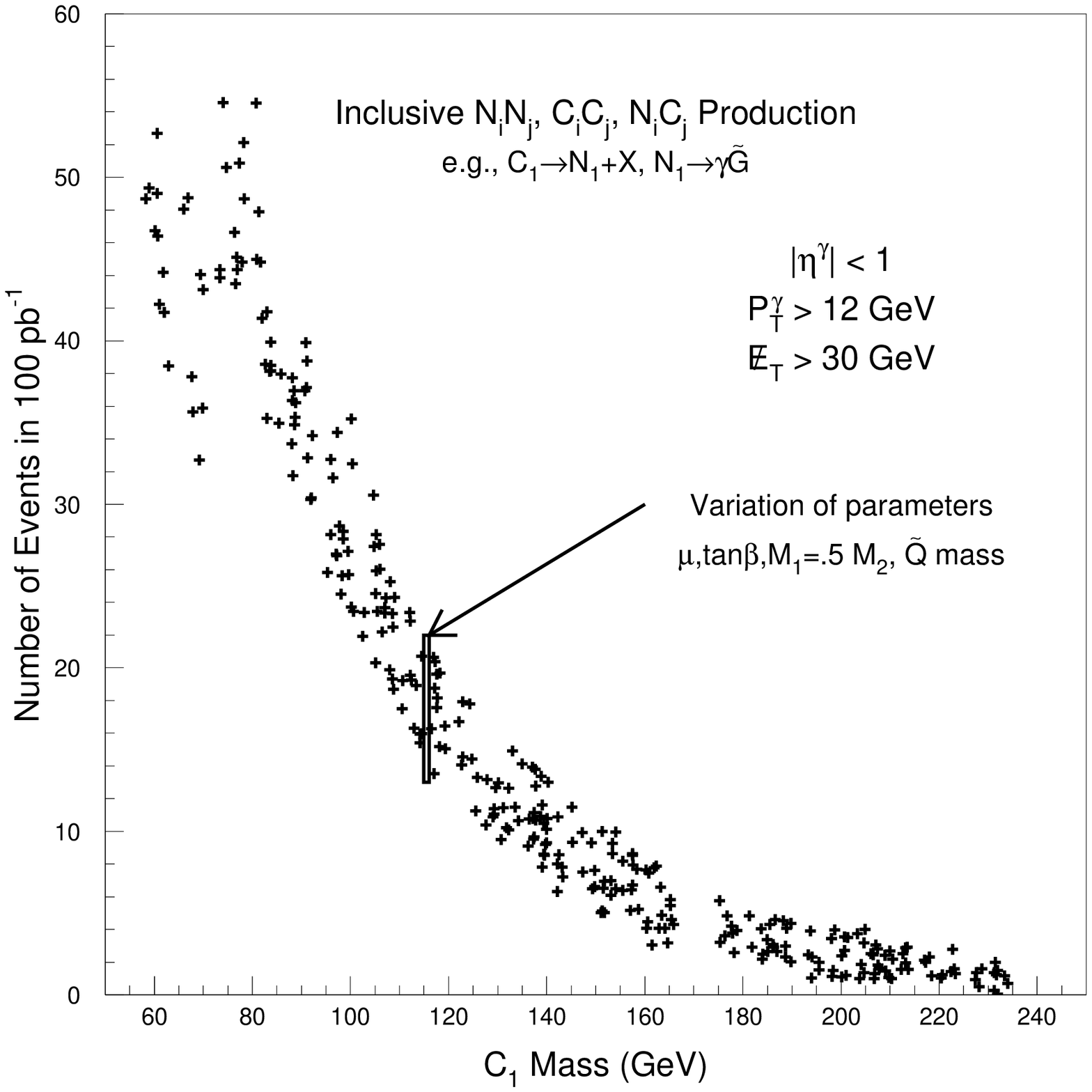}
\caption{The number of $\gamma\gamma\Et +X$ events expected in 100
pb$^{-1}$ of data using the models defined by (\ref{testmodels-eq})
and the cuts explained in the text with $E_T(\gamma)>12$ GeV.
The signal comes from the inclusive production cross sections
for charginos and neutralinos
($\protect{\Ni\Nj}$ and $\protect{\Ni\Cj}$ and $\protect{\Ci\Cj}$)
at the Tevatron with
$\protect\sqrt{s} = 1.8 \; \protect{\rm TeV}$ as a function of the
lightest chargino ${{\CI}}$ mass, assuming gaugino
mass unification.}
\label{figure4}
\end{figure}
\begin{figure}
\centering
\epsfxsize=3.4in
\hspace*{0in}
\epsffile{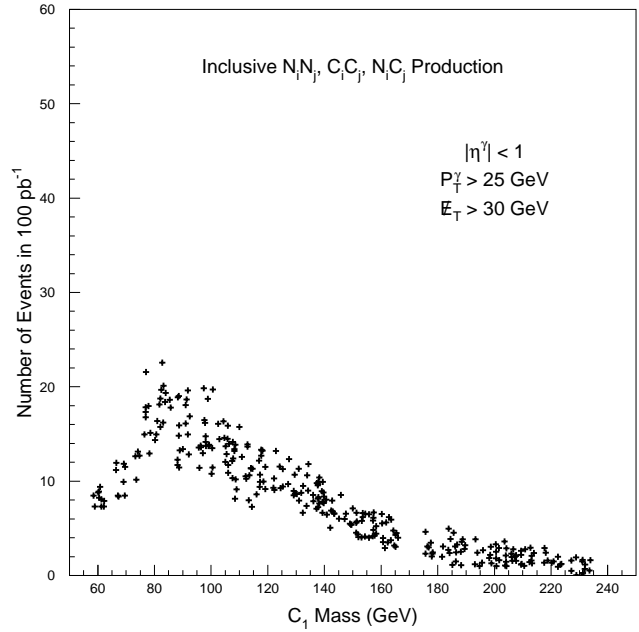}
\caption{As in Fig.~\ref{figure4}, but with $E_T(\gamma) > 25$ GeV.}
\label{figure5}
\end{figure}
vanishes for squark masses above 500 GeV at the Tevatron.
We display the results in terms of the number of expected events
in 100 pb$^{-1}$ as a function of ${{\CI}}$ mass in Figs.~\ref{figure4} and
\ref{figure5}, for
a minimum $E_T(\gamma)$ threshold of 12 and 25 GeV respectively.
The efficiency  for detection of the signal is also displayed in
Fig.~\ref{figure6a} as a function of $m_{\NI}$ and in
Fig.~\ref{figure6b} as a function of $m_{\CI}$ using the
$E_T(\gamma) > 12$ GeV cut.  By comparing
Figs.~\ref{figure4} and \ref{figure5}, we conclude that most
of the photons originating
from models with $m_{ {{\CI}}} \ge 140$ GeV which pass the 12 GeV
$E_T$ cut will also pass the higher threshold.
It is also clear that the
lower $E_T(\gamma)$ threshold substantially increases the
signal for smaller $m_{ {{\CI}}}$
despite the loss of triggering efficiency.

\begin{figure}
\centering
\epsfxsize=3.4in
\hspace*{0in}
\epsffile{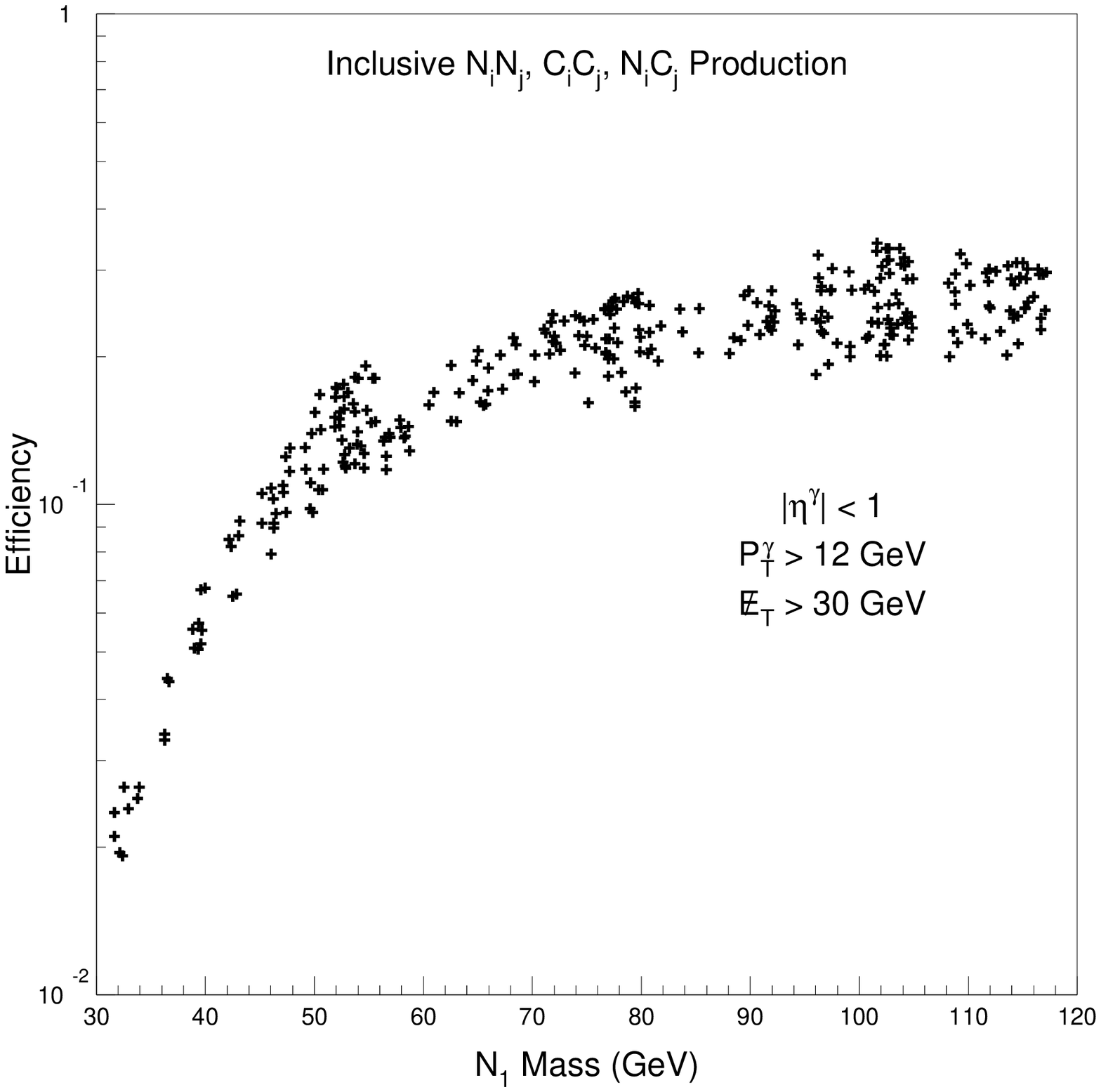}
\caption{The number of generated $\gamma\gamma\Et +X$ events passing cuts
divided by the total for
$E_T(\gamma)>12$ GeV, for the models defined by (\ref{testmodels-eq}).
The signal comes from the inclusive production cross sections
for charginos and neutralinos
($\protect{\Ni\Nj}$ and $\protect{\Ni\Cj}$ and $\protect{\Ci\Cj}$)
at the Tevatron with
$\protect\sqrt{s} = 1.8 \; \protect{\rm TeV}$ as a function of the
lightest chargino ${{\CI}}$ mass, assuming gaugino
mass unification.}
\label{figure6a}
\end{figure}
\begin{figure}
\centering
\epsfxsize=3.4in
\hspace*{0in}
\epsffile{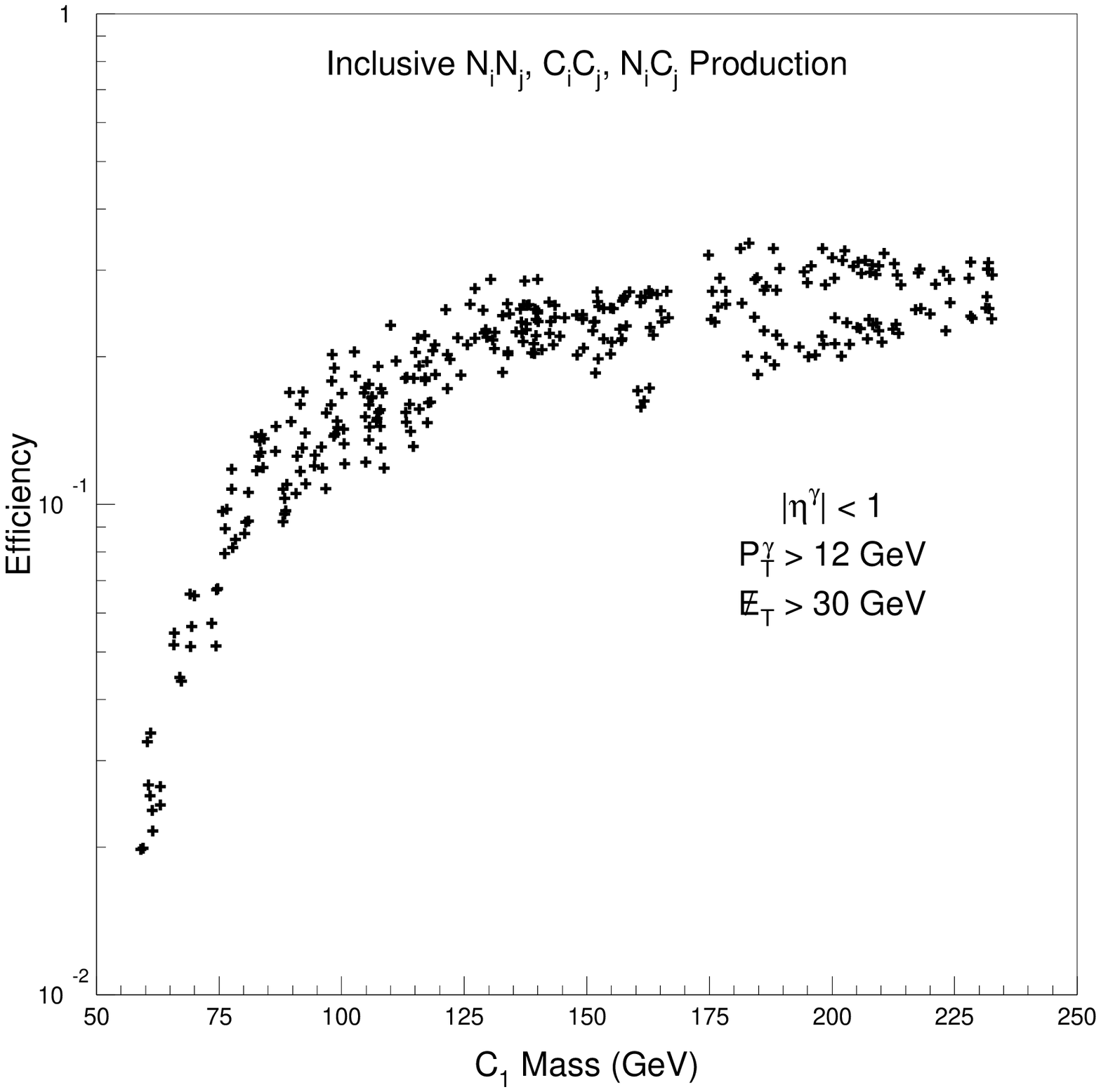}
\caption{As in Figure \ref{figure6a}, but as a function of
$\protect{m_{\CI}}$.}
\label{figure6b}
\end{figure}
\begin{figure}[t]
\centering
\epsfxsize=4in
\hspace*{0in}
\epsffile{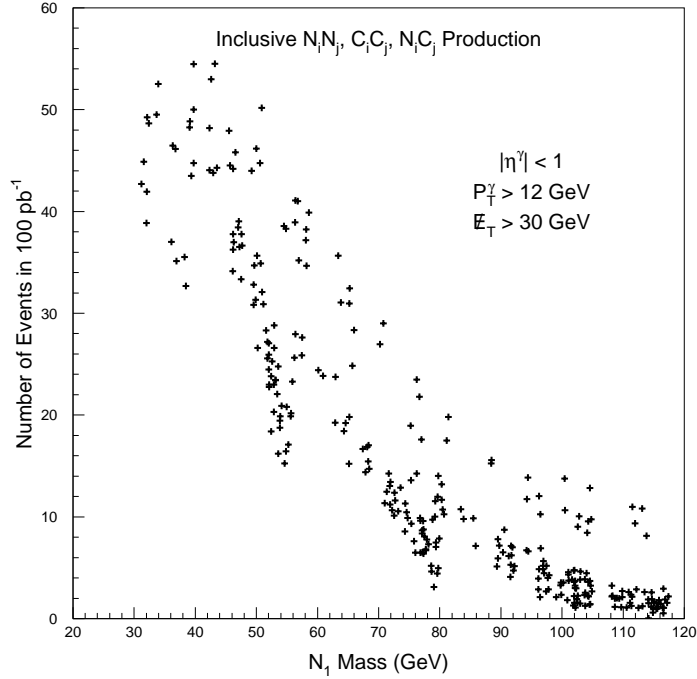}
\caption{The number of $\gamma\gamma\Et +X$ events expected in 100
pb$^{-1}$ of data using the cuts explained in the text with
$E_T(\gamma)>12$ GeV.
The signal comes from the inclusive production cross
sections
for charginos and neutralinos
($\protect{\Ni\Nj}$ and $\protect{\Ni\Cj}$ and $\protect{\Ci\Cj}$)
at the Tevatron with
$\protect\sqrt{s} = 1.8 \; \protect{\rm TeV}$ as a function of the
lightest neutralino ${\NI}$ mass, assuming gaugino
mass unification, and using
the models defined by (\ref{testmodels-eq}).}
\label{figure7}
\end{figure}

These figures suggest that the non--observation of a signal excludes
$m_{\CI} < $ 125 GeV when (\ref{gauginomassunification}) is assumed,
which is well above the pair production
threshold of any LEP upgrade.
The same information, but plotted as a function of the
${\NI}$ mass, is shown in Fig.~\ref{figure7}. From this plot,
we conclude that
a lightest neutralino mass below 70 GeV is excludable in the same manner,
when (\ref{gauginomassunification}) is assumed.
We have not attempted a completely general study of efficiencies
when the gaugino mass unification is not assumed. However, we
do not see any reason to expect significantly lower efficiencies
in the completely general case. In particular, small mass differences
between charginos and neutralinos should have little effect on
the efficiency (for fixed $m_{\NI}$) since the photon energies and
$\Et$, which primarily
determine the signal, depend on the mass and boost of $\NI$.
Therefore by considering Figs.~\ref{figure6a} and \ref{figure6b}
and using the fact that the minimum production cross-section as
a function of $m_{\CI}$ is bounded from below as in Fig.~\ref{inofigure2},
we conclude that it should be possible to exclude $m_{\CI} < 100$ GeV
for $m_{\NI}> 50 $ GeV using the present 100 pb$^{-1}$ of Tevatron
data, even without assuming (\ref{gauginomassunification}).

\begin{figure}[t]
\centering
\epsfxsize=4in
\hspace*{0in}
\epsffile{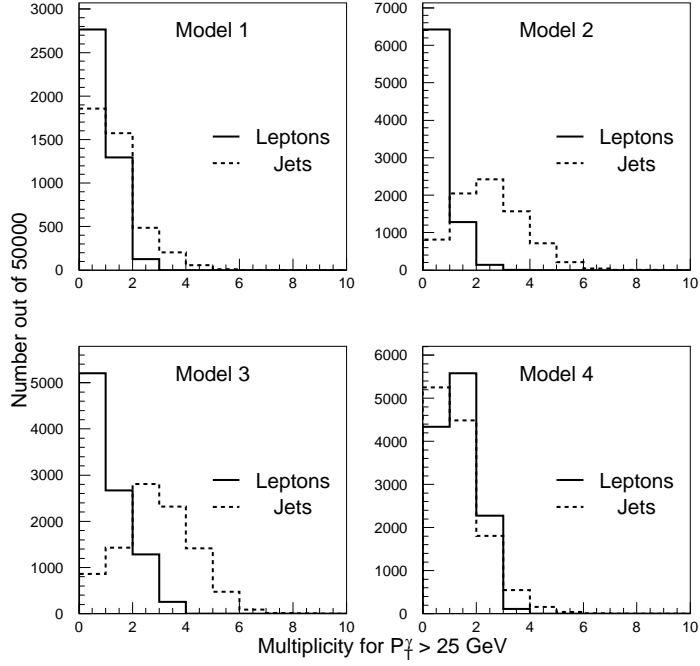}
\caption{The lepton and jet multiplicities for the four models
explained in the text.}
\label{figure8}
\end{figure}

As mentioned previously, the lepton and jet multiplicities of such
events can be large, although they can be sharply reduced
from naive expectations because of limited
detector acceptance, jet definition, and isolation criteria.
This is particularly relevant when the mass splittings among
charginos and neutralinos are relatively small.
Jets $(j)$ are defined using
a standard clustering algorithm with $R$=0.5 and $E_T^{j}>$ 15 GeV,
$|\eta^{j}|<$ 2.5.  The particle energies are smeared using typical
CDF energy resolutions.  Electrons and muons must have
$E_T^{(e,\mu)}>$ 20 GeV
and $|\eta^{(e,\mu)}|<$ 2.0, while being isolated from excess transverse
energy.
We illustrate typical jet and lepton multiplicities for four
specific models in Fig.~\ref{figure8}.  The model parameters are:

$\bullet$ {\bf Model 1}: $M_2 = 100$ GeV, $\mu = -216$ GeV, $\tan\beta = 2.5$,
$m_{\squark} = m_{\slepton} = 1000$ GeV.
One then finds
$m_{\Ni} = (53, 108, 227, 238)$ GeV,
$m_{\Ci} = (108, 240)$ GeV,
$\kappa_{1\gamma} = 0.85.$

$\bullet$ {\bf Model 2}: $M_2 = 150$ GeV, $\mu = -125$ GeV, $\tan\beta = 1.7$,
$m_{\squark} = m_{\slepton} = 1000$ GeV.
One then finds
$m_{\Ni} = (80, 119, 153, 179)$ GeV,
$m_{\Ci} = (134, 182)$ GeV,
$\kappa_{1\gamma} = 0.81.$

$\bullet$ {\bf Model 3}: $M_2 = 200$ GeV, $\mu = 600$ GeV, $\tan\beta = 3.0$,
$m_{\lR} = 150$ GeV, $m_{\snu} = 250$ GeV, $m_{\lL} = 260$ GeV,
$m_{\squark} = 700$ GeV.
One then finds
$m_{\Ni} = (98, 190, 602, 615)$ GeV and
$m_{\Ci} = (189, 614)$ GeV and
$\kappa_{1\gamma} = 0.72.$

$\bullet$ {\bf Model 4}: $M_2 = 225$ GeV, $\mu = 300$ GeV, $\tan\beta = 1.5$,
$m_{\lR} = 105$ GeV, $m_{\snu} = 115$ GeV, $m_{\lL} = 125$ GeV,
$m_{\squark} = 300$ GeV.
One then finds
$m_{\Ni} = (101, 183, 301, 355)$ GeV and
$m_{\Ci} = (176, 350)$ GeV and
$\kappa_{1\gamma} = 0.56.$

We note in passing that Model~4 may be of particular interest,
since it has some
general properties consistent with an alternative
candidate for the CDF $ee\gamma\gamma \Et$ event through ${{\CI}} {{\CI}}$
production.
In this model,
the lepton multiplicity is peaked at 1 but there is also a substantial
component with lepton multiplicity 2 (because of allowed 2-body decays
of $\CI$ and $\NII$ to slepton + $\NI$).
One expects about three chargino/neutralino
events after cuts from this model in the current data taken at the Tevatron. We
will remark further on the chargino pair production
interpretation of the CDF event below.
This model also has light sleptons, so it could have produced
the event through selectron pair production, but the kinematics do
not favor this interpretation because the leptons would be too soft.
Fig.~\ref{figure8} shows how the relative multiplicities
could help distinguish models if a signal is established.

\subsection*{B. Sleptons}
\indent

In most theoretical models, scalar (mass)$^2$ parameters receive
positive contributions proportional to $\alpha_i^2$ and/or $\alpha_i$,
where $\alpha_i$ ($i=1,2,3$) are the gauge couplings felt by the scalar.
Therefore one expects that sleptons with the same $SU(2)_L\times U(1)_Y$
quantum numbers should be degenerate in mass, and should all be
considerably lighter than squarks, with
$m_{\lR} < m_{\sneutrino} < m_{\lL}$ the most plausible mass
ordering. It is therefore interesting to consider slepton discovery
signals at the Tevatron; a corresponding study in the neutralino
LSP scenario appears in \cite{sleptonstev}.
\begin{figure}[t]
\centering
\epsfxsize=4in
\hspace*{0in}
\epsffile{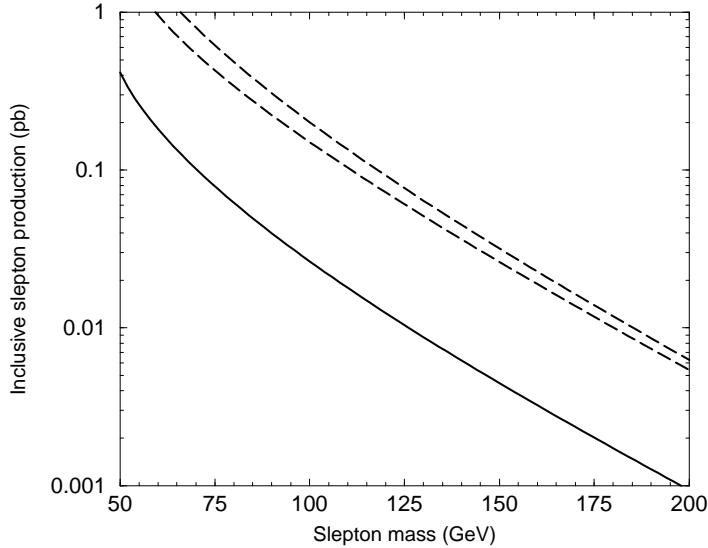}
\caption{Total slepton production cross sections
at the Tevatron with $\protect\sqrt{s} = 1.8 \; \protect{\rm TeV}$,
for $\protect\lR\protect\lR$ as a function of $m_{\lR}$ (solid line),
and left-handed sleptons ($\protect\snu\protect\snu$ and
$\protect\lL\protect\snu$ and
$\protect\lL\protect\lL$) as a function of $m_{\snu}$ (dashed lines).
In the latter case, the lower (upper) dashed line corresponds to
$\tan\beta = 55$ (1.5).
The cross sections shown are summed over slepton flavors, with
slepton masses taken to be flavor-independent.
}
\label{sleptonfigure}
\end{figure}
In Fig.~\ref{sleptonfigure},
we show total Tevatron cross-sections for $\lR\lR$  production
summed over three families
(solid line)
as a function of $m_{\lR}$. The signal for $\ser\ser$ production
with the decay $\ser \rightarrow e\NI$
is $\eegg \Et$, providing a viable candidate
for the single observed CDF event of this type \cite{Event}.
As is
known from the analyses of \cite{Event,DDRT,AKKMM}, such events do not
seem to
have a probable SM interpretation.
In the same figure we show as a function of $m_{\sneutrino}$
the total cross-section for $\sneutrino\sneutrino$,
$\lL\sneutrino$, and $\lL \lL$ production, for
$\tan\beta = 1.5$ and $55$ (dashed lines). Since the
masses of $\sneutrino$ and $\lL$ are related by the sum rule
(\ref{sumrule}), the rates for $\lL\sneutrino$ and $\lL\lL$
production decrease monotonically with larger $\tan\beta$
(for a fixed value of $m_{\sneutrino}$).
The $\lL \snu$ component of the signal is always the largest.
The final states from $\lL \lL$ and $\lL\sneutrino$ production
will depend specifically on the slepton and ${\Ci}$, ${\Ni}$ masses,
but can contain $\gamma\gamma\Et$ and up to three charged leptons.
\begin{figure}[t]
\centering
\epsfxsize=4in
\hspace*{0in}
\epsffile{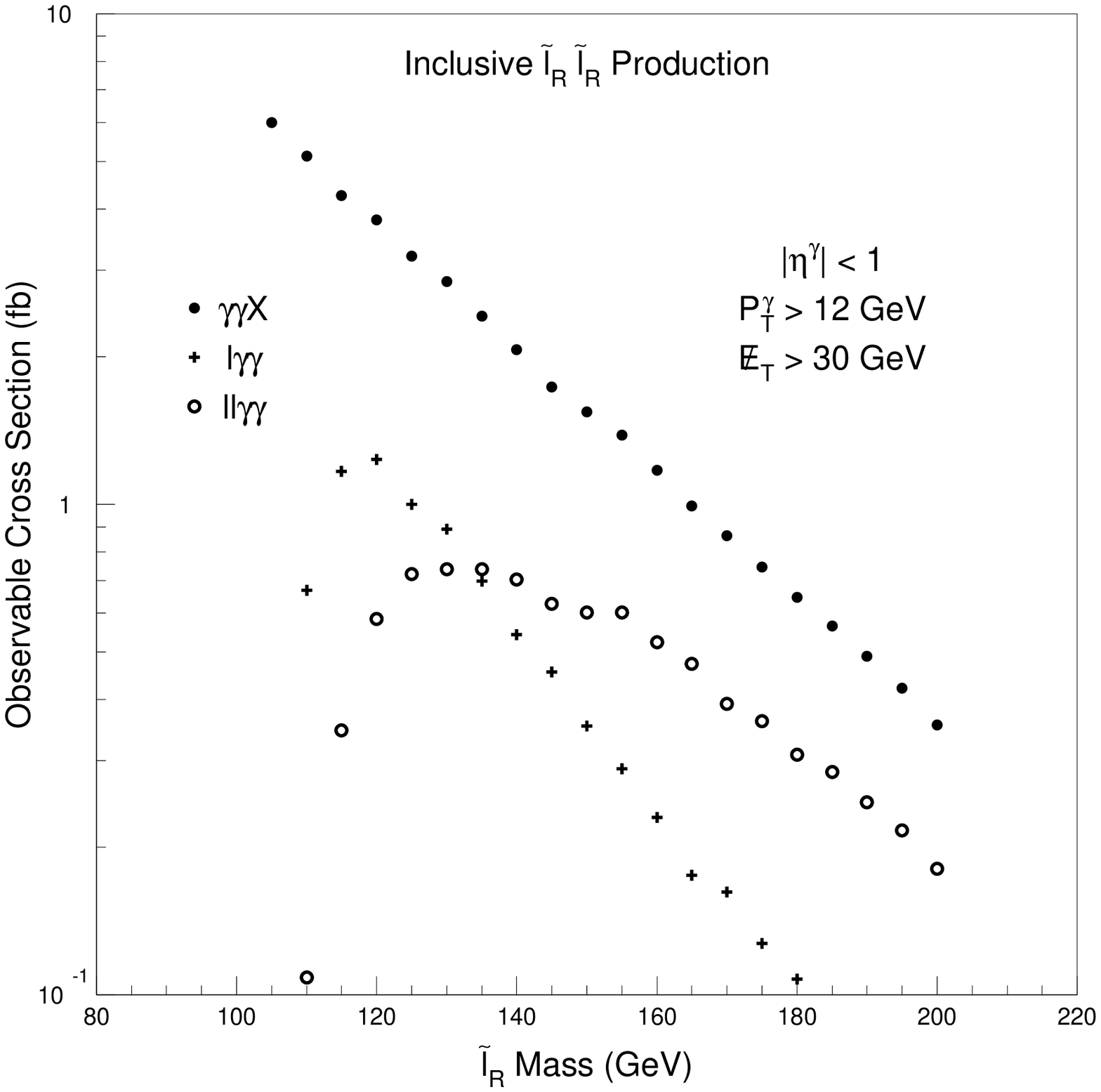}
\caption{Expected $\lR\lR$ signals (including all three lepton flavors)
from Model 4 at the Tevatron with
$\protect\sqrt{s} = 1.8 \; \protect{\rm TeV}$, as a function of
$m_{\lR}$. The total $\gamma\gamma \Et +X$ signal is shown by
the filled circles, while the single lepton and dilepton components
are denoted by crosses and open circles, respectively.}
\label{figure12a}
\end{figure}

\begin{figure}[t]
\centering
\epsfxsize=4in
\hspace*{0in}
\epsffile{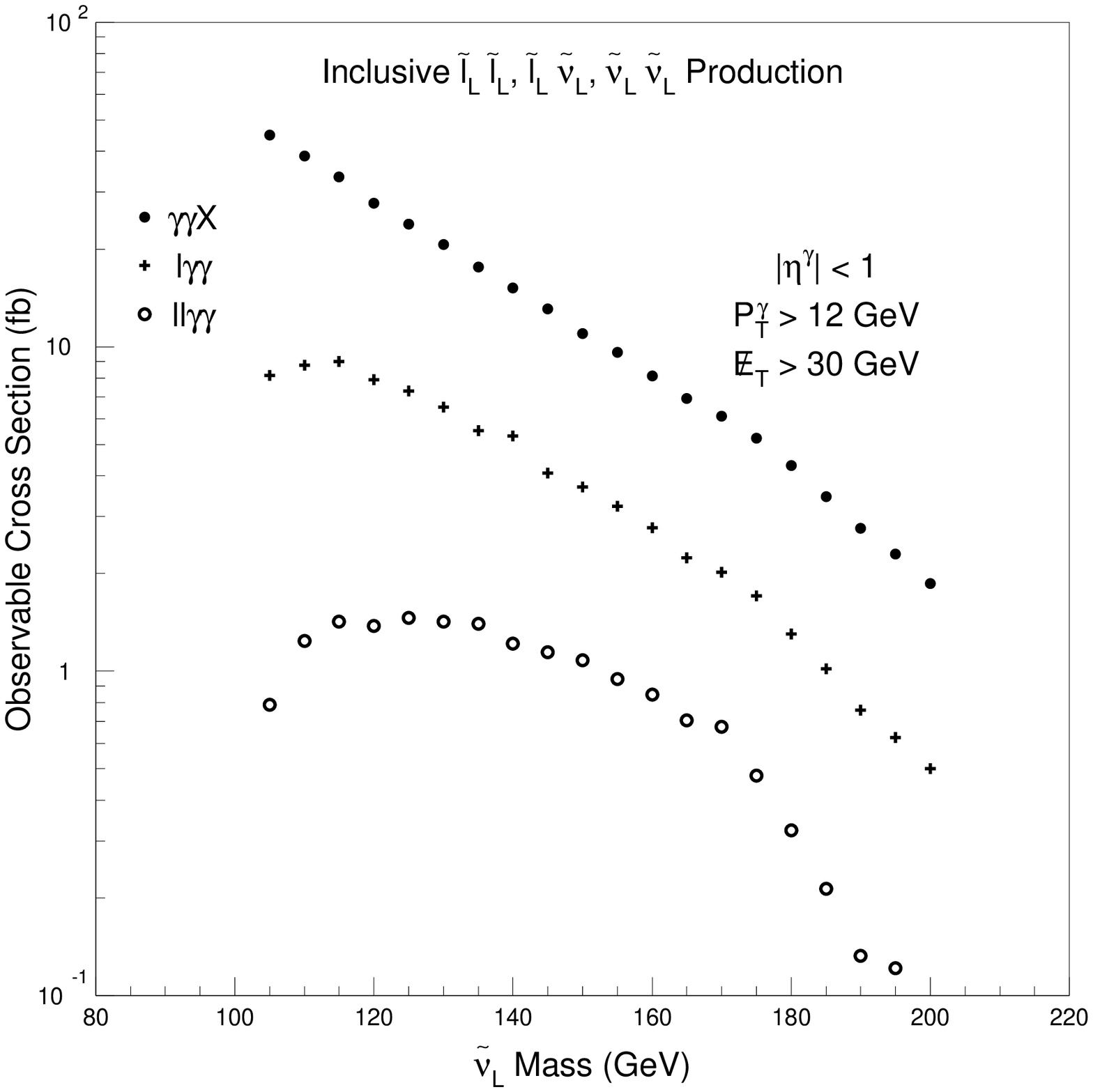}
\caption{Total
expected $\lL\lL$ and $\lL\snu$ and $\snu\snu$
signals (including all three lepton flavors)
from Model 4 at the Tevatron with
$\protect\sqrt{s} = 1.8 \; \protect{\rm TeV}$, as a function of
$m_{\snu}$. The total $\gamma\gamma \Et +X$ signal is shown by
the filled circles, while the single lepton and dilepton components
are denoted by crosses and open circles, respectively.}
\label{figure12b}
\end{figure}

Rather than conduct an extensive survey of slepton signatures,
we consider as a test case the chargino/neutralino sector of
Model 4 of subsection~A.
For the fixed set of gaugino parameters of that model, we further
vary the right--handed selectron and sneutrino masses over the
ranges 100 GeV $< m_{\lR}, m_{\snu} <$ 200 GeV.
It should be noted that
for much of this range, direct decays of slepton to lepton and $\NI$
should dominate for the chosen model.
Since the signals from right-handed
sleptons and from left-handed sleptons have rather different characteristics,
and because the masses of right- and left-handed
sleptons are {\em a priori} unrelated, we choose to display the
results separately. In Fig.~\ref{figure12a} we show the
cross-sections after cuts arising from right-handed slepton production. The
inclusive $\gamma\gamma\Et$ signal,
with or without additional leptons, is denoted by filled circles.
We note that the
dilepton component of the signal (open circles)
is greater than the single lepton
component (crosses) for $m_{\lR}> 130$ GeV. There is also a significant
component with no leptons passing the cuts. In Fig.~\ref{figure12b}
we likewise show the total rate after cuts expected from
$\lL\lL$ and $\lL\snu$, and $\snu\snu$ production, as a
function of $m_{\snu}$. For any
given model, the expected number of events with zero or one lepton
far exceeds the number with two leptons. This is partly because
of the comparatively larger cross-section for $\snu\lL$ and $\snu\snu$
production, but also because some leptons do not pass cuts.

\subsection*{C. Light stop squarks}
\indent

\begin{figure}[t]
\centering
\epsfxsize=4in
\hspace*{0in}
\epsffile{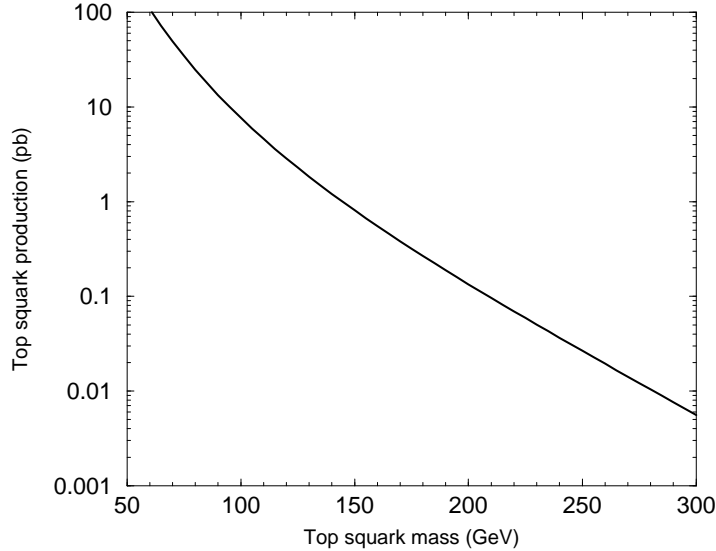}
\caption{Cross section for pair-production of the lighter top squark mass
eigenstate at the Tevatron with
$\protect\sqrt{s} = 1.8 \; \protect{\rm TeV}$.}
\label{topsquarkfigure}
\end{figure}

In specific models, one stop squark mass eigenstate (${\tilde t}_1$)
is often found to be
much lighter than all of the other squarks, and can even be lighter
than the top quark.
If $m_{{\tilde t}_1} < 100 $ GeV, chargino/stop
loops might help to explain \cite{RB} the excess of $R_b$ in the LEP data.
However, we have already seen that in the gravitino LSP scenario,
charginos must be far too heavy for this to occur.
Furthermore, a significant bound can be independently placed on a
light stop squark mass in the gravitino LSP scenario
given the integrated
luminosity already obtained at the Tevatron.
In Fig.~\ref{topsquarkfigure},
we show the total ${\tilde t}_1 {\tilde t}_1^*$ pair production
cross-section as a function of ${\tilde t}_1$ mass. In the gravitino LSP
scenario, this process should lead to spectacular signals
$\gamma\gamma \Et$+jets.

\begin{figure}[t]
\centering
\epsfxsize=4in
\hspace*{0in}
\epsffile{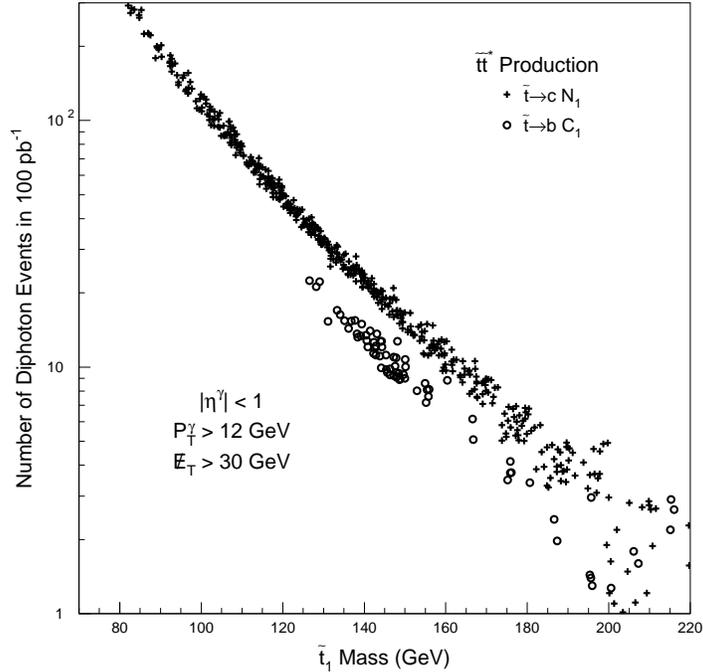}
\caption{Number of $\protect{\gamma\gamma \Et +X}$ events in
100 pb$^{-1}$ as a function of
the lighter top squark mass eigenstate
at the Tevatron with
$\protect\sqrt{s} = 1.8 \; \protect{\rm TeV}$.}
\label{figure10}
\end{figure}
We consider two scenarios, based on the mass orderings $m_{\tilde
t_1}\le m_{\CI} + m_b$ or
$m_{\tilde t_1}>m_{{{\CI}}}+m_b$.  In the first case,
each $\tilde t_1$ cascades through two two--body decays to a
$c\gamma\tilde G$ final state.  In the second, $\tilde t_1$ undergoes
an additional three--body decay to reach a
$bf\bar f\gamma\tilde G$ final state, where $f$ is a fermion.  As
a result, the photons produced in the second case tend to be softer.
For the first case, $\tilde t_1 {\tilde t_1}^{*}$ production
leads to two additional charm jets in the final
state, while $b$--jets and additional leptons or jets are present
for the second.
As before, we ignore such particulars, which could
substantiate a suspected signal, and concentrate
on the same inclusive $\gamma\gamma\Et +X$ signal.
Based on the previous bounds on $m_{{\NI}}$ and $m_{{{\CI}}}$,
we consider models with
$m_{{{\CI}}} +m_b> m_{\tilde t_1} > m_{{\NI}} >$ 70 GeV for the range of
gaugino parameters discussed previously, and a smaller set of models
with $m_{\tilde t_1} > m_{{{\CI}}}+m_b > $ 125 GeV.
The results are illustrated
in Fig.~\ref{figure10}, which
shows the number of expected diphoton events in the present data
sample using the previously defined cuts with $E_T(\gamma )>$12 GeV.
There is a substantially higher detectability of the signal
when $\stoop \rightarrow \CI b$ is not kinematically allowed.
The absence of such events in the present data
sample seems to exclude a ${\tilde t}_1$ lighter than at least 140 GeV
even in the case that $m_{\tilde t_1}>m_{\CI}+ m_b$,
which is already far too heavy to have any effect on the
interpretation of the LEP $R_b$ measurement.
Additionally,
if $m_{\tilde t_1}+m_{{\Ni}} < m_t$, then the decay $t\to {\tilde t_1}
{\Ni}$ can occur, generating $\gamma\gamma\Et$ events
from $t\bar t$ production, but the bounds on $m_{\stoop}$ and
$m_{\NI}$ preclude this.

Note that the limit on the mass of the light stop in the gravitino LSP
scenario is much stronger than for the neutralino LSP scenario in the
case where $\tilde t_1 \to c {\NI}$.  The latter case, which relies on the
signal of two acollinear jets and $\Et$, is limited by the mass splitting
between $\tilde t_1$ and ${\NI}$ which determines the jet $E_T$
spectrum \cite{claes}.
The main limitation of the gravitino LSP scenario is $m_{{\NI}}$ which
sets both the scale of $E_T(\gamma )$ and $\Et$.

\subsection*{D. Other Processes}
\indent

\begin{figure}[t]
\centering
\epsfxsize=4in
\hspace*{0in}
\epsffile{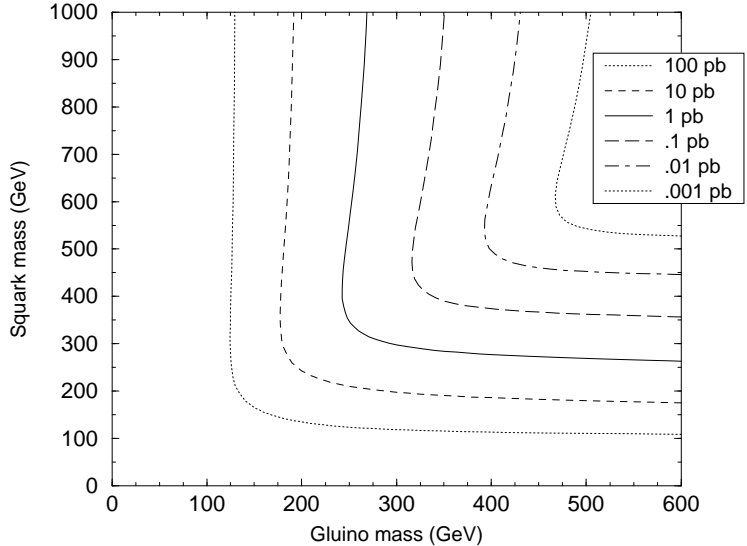}
\caption{Contours of inclusive production cross sections
for gluinos and squarks ($\protect{\gino\gino}$ and $\protect{\gino\squark}$,
$\protect{\squark\squark}$) at the Tevatron with
$\protect\sqrt{s} = 2 \; \protect{\rm TeV}$. All squark flavors are taken
to be degenerate for simplicity.}
\label{gluinosquarkfigure}
\end{figure}

It is a common feature of known
models that the gluino and squarks are quite
heavy. However, the presence of energetic photons in the eventual decay
products means that the detection efficiency is likely to be higher in the
gravitino LSP scenario than in the neutralino LSP scenario.
Therefore it is again interesting to get an idea of the upper limit
on the potential reach of
the Tevatron collider by considering the total inclusive
production of gluinos and squarks
($\gino\gino$, $\gino\squark$, $\squark\squark$).
It must be mentioned that there
are at least two factors which might adversely affect
the efficiency here somewhat, including larger boosts leading to longer
decay lengths for $\NI\rightarrow\gamma\G$ [cf. eq.~(\ref{decleng})],
and losses from photon isolation requirements
with the higher jet multiplicity.
In Fig.~\ref{gluinosquarkfigure},
we show contours of this total production cross
section, in the $(m_{\gino},m_{\squark})$ plane. (For simplicity, we
have assumed degeneracy of all squark flavors.) At least in the case
when gaugino mass unification is assumed, it appears to be doubtful that
gluino pair-production processes can be useful, since the non-observation
of events from chargino and neutralino production together with
(\ref{gauginomassunification}) forces the gluino mass to be too large.
It should be noted that $M_3$ as given by (\ref{gauginomassunification})
is less than the physical pole mass of the gluino
\cite{MartinVaughn} by an amount which is often quite substantial, especially
if the squarks are heavier. This effect makes it even less
likely that processes involving gluino production can compete
with the chargino/neutralino processes already discussed.
The rate for
associated production of ${\Ci}$ or ${\Ni}$ with a gluino or squark
overtakes the ($\gino\gino+\gino\squark+\squark\squark$) production
rate when
$m_{\gino}$ exceeds roughly 400 GeV, depending on the squark masses
\cite{BCKT}.
In this regime, however, even the sum of all
processes involving gluino or squark production should be small
compared to those from chargino and neutralino production, unless
$m_{\squark}$ is significantly less than $m_{\gino}$.
It therefore seems unlikely that gluinos or squarks can be
involved in the discovery process.

\subsection*{E. Comments on the interpretation of the
CDF $ee\gamma\gamma\Et$ event}
\indent

In this section we have emphasized the power of the Tevatron
in setting exclusion limits in the gravitino LSP scenario. Of course,
in this context we must mention that at least one event \cite{Event}
of this general type has been observed at CDF. This event has an energetic
electron and positron, two energetic photons each with $|\eta| <1$,
and large ($>50$ GeV) $\Et$. It easily passes the cuts defining our
signal above.

The most obvious candidate process for this event is selectron pair
production. As has already been discussed in \cite{DDRT,AKKMM},
and recently in some
more detail in \cite{DTW}, one can attempt to explain
the event either in terms of $\ser\ser$ pair production or $\sel\sel$
pair production. From
the kinematic information, we found in \cite{AKKMM} that in either of these two
cases, one has rough bounds
$m_{\tilde e} > 80$ GeV and $38 < m_{\NI} < 100$ GeV.
If one assumes gaugino
mass unification, the lower bound obtained here for $m_{\NI}$ is far
weaker than the lower bound established above from non-observation of chargino
and neutralino events at the Tevatron. As has been
emphasized recently in \cite{DTW}, the energetic electrons in the event
seem to indicate a significant mass difference between $m_{\NI}$ and
$m_{\tilde e}$, in order to have sufficiently energetic electrons
with a high enough probability to explain the event.

Right-handed selectrons have a lower production cross-section
than do left-handed selectrons for a given mass, as can be seen from
Fig.~\ref{sleptonfigure}, and this seems to perhaps favor the idea
that the pair-produced
selectron was left-handed. However, since we are forced to
calculate the probability of this single event using Poisson statistics, this
argument is not on very solid footing.
For example, if the three right-handed
sleptons are degenerate in mass, then the cross-section
to produce {\em any} pair $\lR\lR$ is of course 3 times larger than
the rate for the pair $\ser\ser$ which could explain the event actually
seen. It is not completely clear which of these rates should be taken
in assessing the likelihood of a single observed event.
In any case, the rate before
cuts for $\lR\lR$ production in 100 pb$^{-1}$ is about
1 event for $m_{\lR}= 125$
GeV and $1/2$ event for $m_{\lR} = 145$ GeV. The rates after cuts
are significantly less because of acceptances (see
Fig.~\ref{figure12a}), but
it seems possible that the observed event is an upward
fluctuation in the $\ser\ser$ production process, even with $m_{\ser}-m_{\NI}$
sufficiently large to explain the observed kinematics.

Conversely, although the rates for left-handed selectron production
are larger, one must also note that in this interpretation the expected
number of events with two leptons is always considerably less
than for one or zero leptons (see Fig.~\ref{figure12b}).
The reason for this is that
the production cross-section from $\snu\snu$ and $\snu\lL$ is necessarily
larger than for $\lL\lL$ because of the kinematics dictated by the sum rule
(\ref{sumrule}). It might therefore be viewed as problematic that
the single observed event has two leptons. The limited acceptance for
leptons only exacerbates this problem. Nevertheless, it again seems
to be not entirely out of the question that the event could be
due to $\sel\sel$ production.

 There is another very interesting
possibility, however, illustrated by model 4 above, that the event
could be due to chargino pair-production. We note that depending on parameters,
the chargino pair-production cross section remains
sufficiently large to give 1 event (after cuts) in 100 pb$^{-1}$ up to
at least $m_{\CI}$ = 200 GeV. Now, each produced chargino can decay
into either $\ell\nu\gamma\G$ or $q{\overline q}\gamma\G$.
If the decay is dominantly
through $W$-boson or squark exchange,
then one might expect the $ee\gamma\gamma\Et$
event to be accompanied by many more events with jets or significant
hadronic activity in addition to $\gamma\gamma\Et$. However, this
can be avoided if $\CI$ has kinematically allowed 2-body
decays into left-handed sleptons and not into $W\NI$:
\beq
& &m_{\CI}>m_{\snu}>m_{\NI};\\
& &m_{\CI} - m_{\NI} < m_W .
\eeq
In this case, the chargino should decay as either
$\CI\rightarrow\snu \ell \rightarrow \ell\nu\NI \rightarrow \ell\gamma\Et$
or (if kinematically allowed)
$\CI\rightarrow\lL \nu \rightarrow \ell\nu\NI
\rightarrow \ell\gamma\Et$. Both of these decays have the same signal
(with different kinematics), so that the signal for chargino
pair production will be $\ell^+\ell^{\prime -} \gamma\gamma \Et$
in this case, with a nearly 100\% branching fraction
before cuts. Even though sneutrinos are lighter than charginos, the
chargino production cross-section can be much larger. Of course,
it should be noted that the rate for different flavor leptons in this
case is twice that for like-flavor leptons. Still, we find that it is
possible to obtain kinematics and rates after cuts
which could explain the CDF event. The kinematics of the event together
with the cross-section for chargino pair-production evidently
favor $m_{\CI} \gsim 125 $ GeV in this case.

We should also mention that $\NI\NII$ production can give an $ee
\gamma\gamma\Et$ signal, but it is very difficult to reconcile this
possibility with the observed event, because of the large invariant
mass of the $ee$ pair in the event \cite{Event}.

\section*{4. Supersymmetry with a light gravitino at LEP}
\indent

The LEP collider at CERN will probe
some of the parameter space which is not yet excludable by
the current 100 pb$^{-1}$ data sample collected at the
Tevatron. However, it is important to take into account the
results of Section 3 when assessing the discovery potential of
the various LEP upgrades. At least within the context of gaugino
mass unification, we have found that the lightest neutralino mass can
be bounded from below by 70 GeV, based on the exclusion capability of the
current 100 pb$^{-1}$ data. Similarly, the lighter chargino
mass is bounded below by 125 GeV, and even when (\ref{gauginomassunification})
is not assumed, one has $m_{\CI}> 100$ GeV
for $m_{\NI} > 50 $ GeV.
Therefore, it is immediately clear that one cannot hope
to observe chargino pair production at any of the LEP upgrades
considered here in the gravitino LSP scenario with our assumption that
$\NI\rightarrow\G\gamma$ always
occurs within the detector. Furthermore, the
second lightest neutralino should also not be kinematically accessible at
LEP even in $\NI\NII$ production, at least in the case that gaugino mass
unification (\ref{gauginomassunification}) holds.
The reason for this is that
$m_{\CI} > 125$ (as required by the Tevatron data) and $m_{\NI}< 95$
(as required for accessibility in $e^+ e^-$ collisions with
$\sqrt{s} = 190$ GeV) forces one into a region of parameter
space with rather large $|\mu |$ and gaugino-like $\NI$ and $\NII$,
so that $m_{\NI} + m_{\NII} > $ 210 GeV.
Therefore, it is clear that in the chargino/neutralino sector,
LEP190 can only hope to observe $\NI\NI$ production with signature
$\gamma\gamma\Etot$. Likewise, the existing Tevatron data makes it
impossible for a light stop (or other squark) to be accessible at LEP
with our assumptions. There is a still a
possibility to observe slepton pair
production since, taking into account efficiencies,
the Tevatron cannot set exclusion limits on slepton masses
which are significantly stronger than the indirect one following from
$m_{\ltilde} > m_{\NI} > 70$ GeV. Therefore there is a narrow range
of $\NI$ and slepton masses from no less than 70 GeV up to less than
95 GeV which can be probed at LEP with $\sqrt{s} \leq 190$ GeV.

We begin by considering $\NI\NI$ production in $e^+ e^-$ collisions,
which leads to events with two acoplanar photons and large missing energy.
(A similar study for the NLC has recently been made \cite{SWY}.)
The energy distribution of photons
produced\footnote{We neglect final state interference effects
throughout the following discussion.}
in such events is flat, with endpoints
\beq
& &E_{\rm min}\> < \> E_{\gamone},E_{\gamtwo} \> < \> E_{\rm max}
\label{gammaenergyrange}\\
& &E_{\rm max, min} = {1\over 4}(\sqrt{s} \pm \sqrt{s - 4 m_{\NI}^2}).
\label{endpoints}\eeq
The two photon energies in each event vary over this range
independently, providing a very simple characteristic kinematic signature.
The missing energy in each event is bounded according to
$2 E_{\rm min} < \Etot < 2 E_{\rm max}$ and is peaked at
$E_{\rm beam} \equiv \sqrt{s}/2$.
Two further corollaries are that the distribution of
$E_{\gamma_1}+E_{\gamma_2}$
is the same as that of $\Etot$,
and that the energy distribution of the more (less) energetic photon
observed in each event rises (falls) linearly
with energy. The numerical bounds on photon energies in the $\NI\NI$
signal are, for the various LEP upgrades:
\beq
& &20\>{\rm GeV}\> < \> E_{\gamone},E_{\gamtwo} \> < \> 60\>{\rm GeV}
\qquad (\sqrt{s} = 160 \> {\rm GeV})
\label{rang1}\\
& &18\>{\rm GeV}\> < \> E_{\gamone},E_{\gamtwo} \> < \> 70\>{\rm GeV}
\qquad (\sqrt{s} = 175 \> {\rm GeV})\label{rang2}\\
& &16\>{\rm GeV}\> < \> E_{\gamone},E_{\gamtwo} \> < \> 80\>{\rm GeV}
\qquad (\sqrt{s} = 190 \> {\rm GeV})\label{rang3}
\eeq
for a lower bound $m_{\NI} = 70 $ GeV. For masses nearer threshold,
the range of photon energies of course becomes narrower around
$\sqrt{s}/4$ in each case. Thus the lower bound on
$\NI$ mass from the Tevatron ensures that the $\NI\NI$ signal
at LEP will automatically pass appropriate cuts on soft photons.
This will be useful below in our discussion of cuts and backgrounds.

Several factors affect the production cross-section for $\NI\NI$ at LEP.
Since in the accessible parameter space
$\NI$ is essentially forced to have a large gaugino component,
the s-channel $Z$ boson exchange contribution is suppressed. If sleptons
are light, the diagrams with slepton exchange will dominate.
The diagrams with $\ser$ exchange are usually far more important,
because the $e\ser\NI$ coupling is larger than the $e\sel\NI$ coupling.
The $\NI\NI$ production cross-section is quite sensitive
to the selectron masses, even if the selectrons themselves are not
accessible at LEP. As a result, the discovery reach is always
within a few GeV of the kinematic limit, but for no value of $m_{\NI}$
can one clearly
guarantee discovery at any of the LEP upgrades, because
of low observable rates for large $m_{\ser}$.
At LEP160,
the cross-section is always less than 0.2 pb
for models (with gaugino mass unification) not excludable at the
Tevatron,
and is less than 0.1 pb
for $m_{\NI} > 75$ GeV. These are optimistic upper bounds, and
the cross-sections for less favorable parameters can be much smaller.
This leaves
open only the possibility of perhaps a few events at LEP160
with 25 pb$^{-1}$ per experiment,
for a narrow mass range, optimistically 70 GeV $< m_{\NI} < 77$ GeV.
As we will remark below, there is also a non-trivial background for
such events, so that an unambiguous discovery will require a certain
amount of luck.

\begin{figure}[t]
\centering
\epsfxsize=4in
\hspace*{0in}
\epsffile{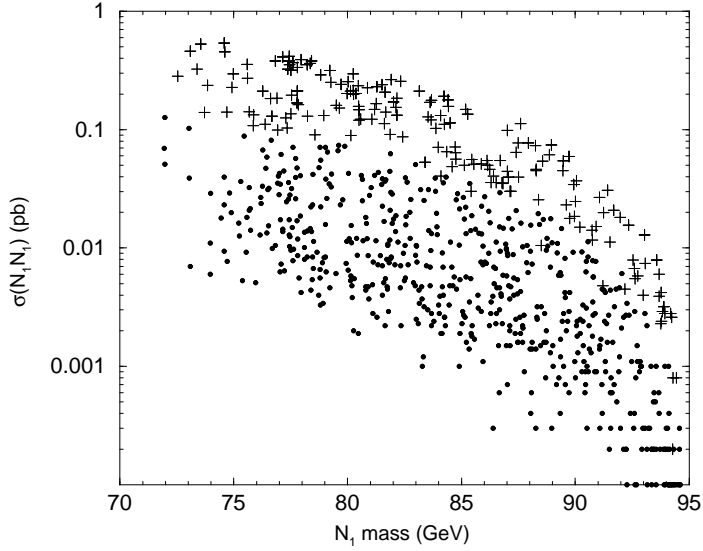}
\caption{Production cross-section for $\protect{\NI\NI}$ at LEP with
$\protect\sqrt{s}= 190$ GeV.
Each point represents a set of model parameters which may not be excludable at
the Tevatron with 100 pb$^{-1}$.
The crosses (dots) represent models with $\protect m_{\ser}$
less (greater) than 175 GeV.}
\label{n1sigmalep}
\end{figure}

The prospects for discovery (or confirmation) are clearly much brighter
at LEP190 with 500 pb$^{-1}$ per experiment, both because of the kinematic
reach and the greater luminosity. In Figure \ref{n1sigmalep}, we show
a scatter plot of the total $\NI\NI$ cross-section at $\sqrt{s} =$ 190 GeV.
Each point on this plot corresponds to a set of model parameters which
plausibly could have avoided detection at the Tevatron with the current
integrated luminosity, based on the results of the previous section.
To illustrate the dominance of the $\ser$ exchange diagrams,
models with $m_{\ser} < 175$ GeV are denoted by crosses, while those
with 175 GeV $<m_{\ser}<$ 500 GeV are denoted by dots.
(The CDF $ee\gamma\gamma\Et$ event could perhaps be explained by
$\ser\ser$ production in models of the former category.)
The gaugino mass unification condition (\ref{gauginomassunification})
is assumed here. We have taken into account initial state radiation
effects which imply
a small ($\lsim10$\%) reduction in the signal; it should be noted
that such effects are larger when $Z$ boson exchange dominates
because of radiative
return to the $Z$ peak. To these cross-sections
one must apply detector cuts, e.g.
\cite{detectorcuts}:
\beq
& & |\cos \theta_\gamma | < 0.95\> ,
\label{anglecut}\\
& & (p_T)_\gamma > 0.065 \> E_{\rm beam} 
{},
\label{ptcut}
\eeq
for each photon. The discovery reach with 500 pb$^{-1}$ extends
up to within a few GeV of the kinematic limit.
Clearly the presence of a light $\ser$ provides much more favorable
discovery prospects.
However,
there is no guarantee of discovery of a light $\NI$ if $m_{\ser}$
is large,
even with this amount of luminosity, and even for the most favorable
kinematics.

We now turn to the question of backgrounds. The
ordinary QED process $e^+ e^- \rightarrow \gamma\gamma $ production
has a large cross section, but can easily be discriminated against
with a cut on missing energy or equivalently
$E_\gamma < 0.8 E_{\rm beam}$ for each photon. The most important
physics backgrounds for the $\gamma\gamma\Etot$ signal come from
$\gamma\gamma\nu_i{\overline\nu}_i$ ($i=e,\mu,\tau$)
with two separately gauge-invariant sets of diagrams:

A) $e^+ e^- \rightarrow \gamma\gamma Z^{(*)}$ with $Z\rightarrow \nu_i
{\overline \nu}_i $. (3 Feynman diagrams);

B) $e^+ e^- \rightarrow \gamma\gamma \nu_e {\overline \nu}_e $ through virtual
$W$-boson exchange. (7 Feynman diagrams).

We have computed these backgrounds using CompHEP
\cite{CompHEP}, a specialized package for automated calculation
of high-energy elementary particle processes, with results
fed into BASES, a Monte Carlo phase-space integration program.
The processes
$e^+ e^- \rightarrow \gamma\gamma \nu_\mu {\overline \nu}_\mu $
and $e^+ e^- \rightarrow \gamma\gamma \nu_\tau {\overline \nu}_\tau $
receive contributions only from the type A diagrams.
At $\sqrt{s} = (160,
175, 190)$, they each contribute $(49, 37, 30)$ fb
to the background for $\gamma\gamma\Etot$ after the
cuts (\ref{anglecut}), (\ref{ptcut}). When the final state
is $\gamma\gamma \nu_e {\overline \nu}_e $, one must take into account
a significant interference between the diagrams of types A and B.
The diagrams of type A clearly dominate in the kinematic regime
characterized by a missing invariant
mass $\minvisible$ very close to $M_Z$. [Here $\minvisible^2$ =
$(p_{e^+} + p_{e^-} - p_{\gamone} - p_{\gamtwo})^2$.]
In that regime, the interference with the type B diagrams
are a small perturbation and in any case only affect 1/3 of the
background.
For slightly larger $\minvisible$, however, the type B diagrams do
have a substantial interference with the off-peak type A diagrams.
The overall effect is one of
constructive interference, but the sign is not
definite for all kinematic configurations.
At $\sqrt{s} = (160, 175, 190)$, we find $(61,49,42)$ fb
for $e^+ e^- \rightarrow \gamma\gamma \nu_e {\overline \nu}_e $
after the cuts (\ref{anglecut}), (\ref{ptcut}).

Since the $\gamma\gamma\nu{\overline \nu}$ backgrounds have larger
support for relatively soft photon energies, one can
reduce them somewhat by imposing the cut
\beq
0.2 < E_\gamma/E_{\rm beam} < 0.8
\label{photonenergycut}
\eeq
on each photon; the upper limit easily eliminates the $e^+ e^-\rightarrow
\gamma\gamma$ process as we have already mentioned. The cut
(\ref{photonenergycut})
has little or no effect on the signal, as can be seen from
(\ref{gammaenergyrange})-(\ref{rang3}).
After imposing this cut in addition to the detector cuts
(\ref{anglecut}), (\ref{ptcut}), we find a remaining background
at $\sqrt{s} = (160,175,190)$ of
$(29,24,21)$ fb from $\gamma\gamma\nu_e{\overline \nu}_e$,
and
$(27,22,18)$ fb from each of $\gamma\gamma\nu_\mu{\overline \nu}_\mu$
and $\gamma\gamma\nu_\tau{\overline \nu}_\tau$.

In order to more strongly
reduce the backgrounds we can impose a cut on the
missing invariant mass of
\beq
10\> {\rm GeV} < \minvisible < 80\> {\rm GeV}\>.
\label{missingmasscut}
\eeq
The upper limit is to avoid the $\gamma\gamma\nu{\overline\nu}$
physics backgrounds, while
the lower limit eliminates a potentially large (several hundred fb)
\begin{figure}[t]
\centering
\epsfxsize=4.in
\hspace*{0in}
\epsffile{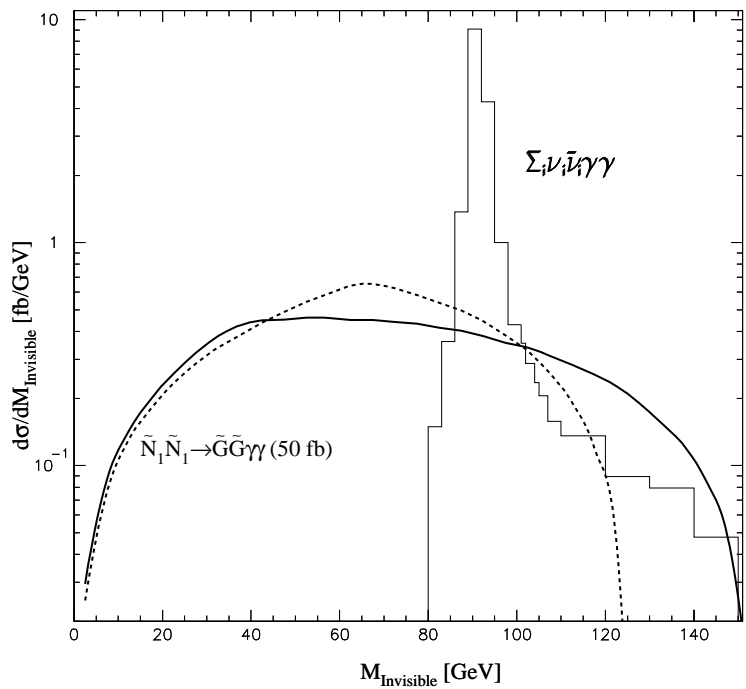}
\caption{Distribution of the missing invariant mass in
$\protect{\gamma\gamma\Etot}$ events at LEP with
$\protect\sqrt{s}= 190$ GeV. Angular and photon energy cuts have been applied
as described in the text.
The lighter solid line is the remaining
total background  (56 fb) for all three
neutrino species. The signals for $\protect{m_{\NI} = 75}$
and 90 GeV are the solid and dashed lines,
respectively, with an arbitrary
choice of 50 fb for the signal before cuts in each case.}
\label{minvsigmalep}
\end{figure}
\begin{figure}[t]
\centering
\epsfxsize=4.in
\hspace*{0in}
\epsffile{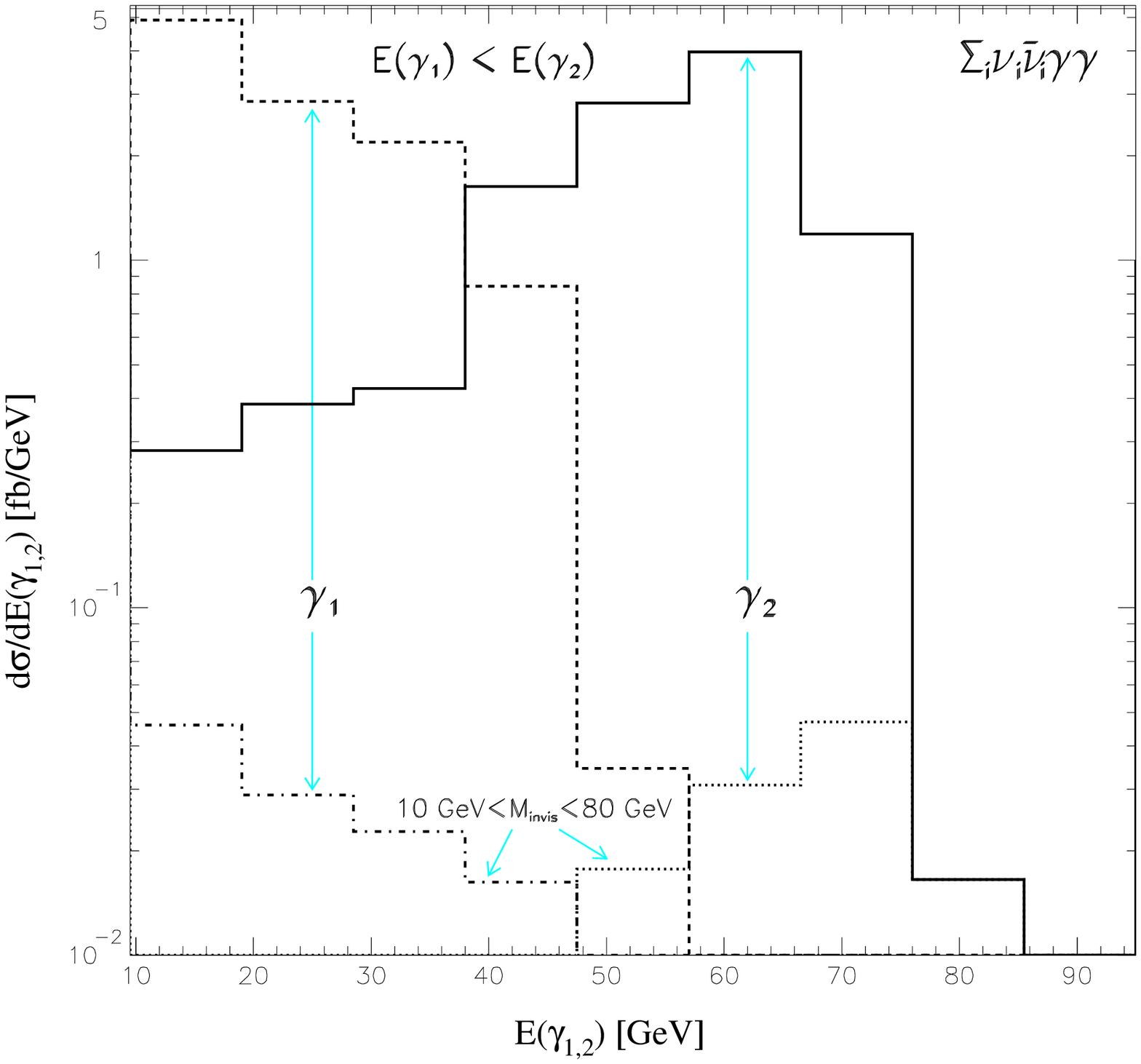}
\caption{Distribution of photon energies for
$\protect{\gamma\gamma\nu{\overline \nu}}$ backgrounds at
LEP with $\protect\sqrt{s}= 190$ GeV.
Detector cuts have been applied as described in the text.
The solid (dashed) line is the distribution for the more (less)
energetic photon $\gamma_2$ ($\gamma_1$)
in each event. The dotted (dot-dash) lines are the same
distributions
after the cut on $\protect{\minvisible}$ described in the text.
}
\label{egammasigmalep}
\end{figure}
detector background
following from $e^+e^-\rightarrow\gamma\gamma(\gamma)$ with one photon
lost in the beam direction or in an insensitive part of the detector.
(The part of this background due to photons lost in the beam direction
is also substantially reduced by imposing a lower bound cut
on $\spt$.) The signal vanishes at the endpoints of the
distribution $\minvisible=0$, $2E_{\rm max}$ and is broadly distributed
in between. The greater part of the signal will always pass all of the
cuts, although a significant part of the
signal will necessarily have to be eliminated by the cut
(\ref{missingmasscut}).
After imposing this cut in addition to
(\ref{anglecut}), (\ref{ptcut}), the total $\gamma\gamma\nu{\overline
\nu}$ backgrounds at $\sqrt{s} = (160,175,190)$ GeV are only $(1.8, 1.3, 1.0)$
fb respectively.
Finally, imposing the cut ({\ref{photonenergycut}) on top of these cuts
reduces the background to a completely negligible level.

The distribution in $\minvisible$
for signals and backgrounds at $\sqrt{s} = $ 190 GeV
are shown in Figure \ref{minvsigmalep}.
In this figure we have arbitrarily chosen a
total signal (before cuts) of 50 fb, with $\NI$ masses of 75
and 90 GeV. The differential
cross-sections shown are after the detector cuts
(\ref{anglecut}), (\ref{ptcut}) and photon energy cuts (\ref{photonenergycut}).
The total $\gamma\gamma\nu{\overline\nu}$
background shown amounts to $56$ fb, but is reduced to a negligible
level by the $\minvisible$ cut.
Note, however, the significant overlap in invariant
missing mass for the backgrounds and the signals.
The signal distribution in $\minvisible$ is broadly
peaked below the 80 GeV cut, and vanishes near $\minvisible = 0$.
We conclude that even in the worst-case kinematic situation, the
efficiency for detecting $\NI\NI$ should exceed 50\% after cuts
at LEP190.
Thus a 40 fb signal before cuts should provide a 10 event
discovery after cuts with 500 pb$^{-1}$. By comparing with Fig.~15,
we conclude that LEP190 should be able to unambiguously observe
$\NI\NI$ production for $m_{\NI}$ up to at least 85 GeV
if $m_{\ser} \lsim 175$
GeV, assuming gaugino mass unification. The exclusion capability
decreases for larger $m_{\ser}$, however.
The $\gamma\gamma\nu{\overline \nu}$ background is more problematic
at $\sqrt{s} = $ 160 GeV with 25 pb$^{-1}$ or
$\sqrt{s} = $ 175 GeV with 10 pb$^{-1}$, where only at most a
few signal events are expected, and the signal distribution in
$\minvisible$ again overlaps with the $Z$-boson peak.

Conversely, Figure \ref{egammasigmalep} shows the distributions for photon
energies at $\sqrt{s} = 190$ GeV for the backgrounds,
before and after the $\minvisible$ cut. All of the
$\gamma\gamma\nu{\overline\nu}$ contributions have been included here.
The two distributions correspond to the more and less energetic photon in
each background event, after the detector cuts (\ref{anglecut}), (\ref{ptcut}).
 After imposing in addition the cuts
(\ref{photonenergycut}) and (\ref{missingmasscut}), the total
$\gamma\gamma\nu{\overline\nu}$ background is reduced
to a fraction of a femtobarn. The signal from $\protect{\NI\NI}$ production
(not shown)
is characterized by a linearly rising (falling) distribution
for the more (less) energetic photon in each event, with endpoints
$\protect{E_{\rm min}}$ and $\protect{E_{\rm max}}$ as given above.
(Note the log scale in Fig.~\ref{egammasigmalep}.)

We now turn to the question of slepton pair production signals at LEP.
In general, slepton masses up to within a few GeV of the kinematic
limit should lead to visible signals with 500 pb$^{-1}$ at LEP190. If
right-handed
sleptons are kinematically accessible, one finds that the cross-section
for $\ser\ser$ production is generally somewhat larger than those for each
of $\smur\smur$ and $\staur\staur$, because of the positive contribution
of diagrams with t-channel exchange of gaugino-like $\NI$.
The pair production of electron sneutrinos can be very strongly
suppressed, because of destructive interference
from chargino exchange, even with $\CI$ required to be heavier
than 125 GeV. Fortunately, pair production of muon and tau
sneutrinos does not suffer this suppression, and those cross-sections are
always large up to within a GeV or two of the kinematic limit.
Because of the sum rule (\ref{sumrule}),
it seems quite unlikely that pair production
of left-handed selectrons can be a discovery process at LEP in the
gravitino LSP scenario considered in this paper.
However, the cross-section for
$\sel\ser$ can be even larger than for $\ser\ser$ production when both are
kinematically accessible,
because of a large contribution from exchange of gaugino-like neutralinos.

It is important to note that for the (quite narrow) range of
masses which are
accessible at LEP and which cannot already be ruled out at the Tevatron,
each slepton has one (and only one) allowed 2-body decay mode,
namely $\ltilde \rightarrow \NI \ell$. This decay is never strongly
suppressed because $\NI$ always has a significant gaugino component.
Therefore,
charged slepton production will essentially always give rise to the signal
$\ell^+\ell^-\gamma\gamma\Etot$,
while sneutrino production, like $\NI\NI$ production,
can give rise only to $\gamma\gamma\Etot$.
The leptons appearing in $\gamma\gamma\ell^+\ell^-\Etot$ events from
charged slepton pair production at LEP
should necessarily be quite soft, because there cannot be a large
mass difference between the slepton and $\NI$ and the sleptons cannot
have a large boost. However, the SM
backgrounds for such processes are extremely small. Taking into account
the cuts (\ref{anglecut}), (\ref{ptcut}), one finds that
$\gamma\gamma ZZ$ production is always below threshold at
LEP160, LEP175, and LEP190, while $\gamma\gamma WW$ is only above
threshold at LEP190. Using CompHEP we have found that the latter
process only contributes about $0.1$ fb to the
$\ell\ell\gamma\gamma\Etot$ background at $\sqrt{s}=190$ GeV. There
is also a background for $ee\gamma\gamma\Etot$ and $\mu\mu\gamma\gamma
\Etot$
from the process $e^+e^- \rightarrow Z^{(*)}\gamma\gamma$ with
$Z\rightarrow\tau^+\tau^-$
and leptonic $\tau$ decays, but this is very small.
Similarly, the photons produced in sneutrino pair production should
be softer than those found in $\NI\NI$ events. Since slepton interactions
are not expected to exhibit significant flavor violation, we can conclude
by noting that the signatures for the gravitino LSP scenario
at LEP are always
$\gamma\gamma\Etot$ (from $\NI\NI$ and $\snu\snu$ production)
and $e^+e^-\gamma\gamma\Etot$, $\mu^+\mu^-\gamma\gamma\Etot$, and
$\tau^+\tau^-\gamma\gamma\Etot$ (from charged slepton production).

\section*{5. Concluding Remarks}
\indent

In this paper, we have studied discovery signals for supersymmetry
with a gravitino LSP at the Tevatron and at LEP. If the decay
$\NI\rightarrow\gamma\G$ occurs
within the detector, then supersymmetric
phenomenology at colliders
will have a very bright future. Indeed, the existing Tevatron data of 100
pb$^{-1}$  should allow the exclusions
$m_{\NI}>70$ GeV and $m_{\CI} > 125$ GeV at least in models
obeying the gaugino mass unification condition (\ref{gauginomassunification}).
For $m_{\NI}> 50$ GeV, it should be possible to exclude $m_{\CI} < 100$ GeV in
a model-independent way.
The reach is much higher.
These results rely on the fact that every supersymmetric event
contains two potentially detectable energetic photons and $\Et$,
yielding a high (up to 30\%) detection efficiency.
We emphasize that this efficiency is not expected to be significantly
reduced by small mass splittings between charginos and neutralinos,
since both the photon energies and the $\Et$ depend only on the
mass and boost of the $\NI$. If the single CDF $ee\gamma\gamma\Et$
event is an example of such an event, then it is not unlikely that
an upgraded Tevatron with $\sqrt{s} = 2$ TeV and
$\geq 2$ fb$^{-1}$ of data can establish a discovery. In any case,
Tevatron upgrades will continue to make strong inroads into the parameter
space of the gravitino LSP scenario. The reach and exclusion capability
can be estimated for Tevatron upgrades using Figs.~\ref{inofigure1}
and \ref{inofigure2}
and assuming efficiencies $\gsim 15$\% as found in
Figs.~\ref{figure6a} and \ref{figure6b}. (Note that the increase in
$\sqrt{s}$ from 1.8 TeV to 2 TeV makes such estimates conservative.)

It is quite possible that LEP190 with 500 pb$^{-1}$ of data can make
the discovery if the lightest neutralino is kinematically accessible.
At least in the
case of models not already excludable by the Tevatron
with gaugino mass unification, only $\NI\NI$ and
slepton pair-production can be explored at LEP, with possible signals
$\gamma\gamma\Etot$ and $\ell^+\ell^-\gamma\gamma\Etot$. We found that
appropriate cuts on the missing invariant mass and on photon energies
can reduce the $\gamma\gamma\Etot$
backgrounds to a negligible level while keeping intact
at least 50\% of the signal even in the worst kinematic situation.
The discovery reach extends to within a few GeV of the kinematic limit.
An important factor in the
$e^+e^- \rightarrow \NI\NI \rightarrow \gamma\gamma \Etot$
search is the mass of the right-handed
selectron. If $m_{\ser} \lsim 175$ GeV, then there should be at
least 10 events after cuts in 500 pb$^{-1}$ at $\sqrt{s} = 190$ GeV
for $m_{\NI}\lsim 85$ GeV, but the rate can be much lower for larger
$m_{\ser}$. In this sense, any exclusion limits will be dependent
on assumed upper bounds for $m_{\ser}$. If $\ser$ is light, then
LEP160 and LEP175 can observe a few events.

Although we have not studied future colliders here,
it seems clear that both the Large Hadron
Collider and Next Linear Collider will be very effective discovery machines if
the detectors have
good efficiency for detecting isolated energetic photons and $\Et$.
If the reported $ee\gamma\gamma\Et$ CDF event is interpreted as slepton
or chargino production, it seems essentially certain that the NLC
will detect supersymmetric events, and that the LHC also will if the
detectors are sufficiently good with photons and $\Et$.

In general, the ability of the present Tevatron data sample to bound the
gravitino LSP scenario emphasizes the importance of
photon detection.  This component should not be ignored
in future detector design.  Also, it would be useful to
have photon pointing information, in case the $\NI\rightarrow\gamma\G$
decay length is macroscopic. As we mentioned in section 2, it is certainly
possible that this is so, leading to more single photon events than
diphoton events. In that case, one can imagine discovering supersymmetry
using the usual well-known discovery signals of the neutralino LSP
scenario, supplemented by a fraction of these events with one additional
energetic photon. Measuring this fraction would provide a powerful
piece of information in disentangling the signal, especially if it
can be combined with measurements of the $\NI$ decay length.

The gravitino LSP possibility also provides a rich area for theoretical
explorations. This scenario necessarily implies a low scale of
spontaneous supersymmetry breaking, but the precise mechanism for this
breaking and its communication to the fields of the MSSM remain open
questions \cite{GaugeMediated}. The connections between
supersymmetry breaking and the $\mu$ parameter and other aspects of
electroweak symmetry breaking are also interesting questions
\cite{DGP}. We should also mention that if the gravitino is the LSP,
then the lightest neutralino is of course no longer a cold dark matter
candidate. It remains to be seen if one can obtain a viable dark
matter scenario; for a recent proposal see \cite{BMY}. It
seems clear that such issues merit further investigation.

\section*{Acknowledgments}
\indent

We are indebted to Bob Blair, Marty Einhorn, Henry Frisch,
Tony Gherghetta, Joey Huston, and Eve Kovacs
for helpful discussions. This work was supported in part by
the U.S. Department of Energy.



\begin{thebibliography}{99}
\singlespaced

\bibitem{Fayet}
    P.~Fayet, \PLB{70}{1977}{461}; \PLB{86}{1979}{272};
    \PLB{175}{1986}{471}; and in ``Unification of the fundamental particle
interactions", eds.~S.~Ferrara, J.~Ellis,   P.~van Nieuwenhuizen
(Plenum, New York, 1980) p.~587.

\bibitem{oldmodels}
    M.~Dine, W.~Fischler, M.Srednicki, \NPB{189}{1981}{575};
    S.~Dimopoulos, S.~Raby, \NPB{192}{1981}{353};
    M.~Dine, W.~Fischler, \PLB{110}{1982}{227};
    M.~Dine,  M.~Srednicki, \NPB{202}{1982}{238};
    L.~Alvarez-Gaum\'e, M.~Claudson, M.~Wise, \NPB{207}{1982}{96};
    C.~Nappi, B.~Ovrut, \PLB{113}{1982}{175};
    M.~Dine, W.~Fischler, \NPB{204}{1982}{346};
    S.~Dimopoulos, S.~Raby, \NPB{219}{1983}{479}.

\bibitem{GaugeMediated}
    For recent examples see M.~Dine, A.~E.~Nelson, \PRD{48}{1993}{1277};
    M.~Dine, A.~E.~Nelson, Y.~Shirman, \PRD{51}{1995}{1362};
    M.~Dine, A.~E.~Nelson, Y.~Nir, Y.~Shirman, \PRD{53}{1996}{2658}.

\bibitem{Raxion}
    A.~E.~Nelson, N.~Seiberg, \NPB{416}{1994}{46};
    J.~Bagger, E.~Poppitz, L.~Randall, \NPB{426}{1994}{3}.

\bibitem{BhattacharyaRoy}
    T.~Bhattacharya, P.~Roy, \PLB{206}{1988}{655};
    \NPB{328}{1989}{481}.

\bibitem{cosmoconstraints}
    H.~Pagels, J.~R.~Primack, \PRL{48}{1982}{223};
    T.~Moroi, H.~Murayama, M.~Yamaguchi, \PLB{303}{1993}{289}.

\bibitem{review}
    For a review, see H.~Baer et al., ``Low-energy
    supersymmetry phenomenology", hep-ph/9503479.

\bibitem{decay}
    N.~Cabibbo, G.~R.~Farrar, and L.~Maiani, \PLB{105}{1981}{155};
    M.~K.~Gaillard, L.~Hall, I.~Hinchliffe, \PLB{116}{1982}{279};
    J.~Ellis, J.~S.~Hagelin, \PLB{122}{1983}{303};
    D.~A.~Dicus, S.~Nandi, and J.~Woodside, \PLB{258}{1991}{231}.

\bibitem{SWY}
    D.~R.~Stump, M.~Wiest, C.~P.~Yuan, preprint hep-ph/9601362.

\bibitem{KNW}
    K.~Kiers, J.~N.~Ng, G.~Wu, preprint hep-ph/9604338.

\bibitem{DTW}
    S.~Dimopoulos, S.~Thomas, J.~D.~Wells, preprint hep-ph/9604452.

\bibitem{Event}
    S.~Park, ``Search for New Phenomena in CDF'', $10^{\rm th}$ Topical
    Workshop on Proton--Anti-proton Collider Physics, edited by
    Rajendran Raja and John Yoh, AIP Press, 1995.

\bibitem{DDRT}
    S.~Dimopoulos, M.~Dine, S.~Raby, S.~Thomas, \PRL{76}{1996}{3494}.

\bibitem{AKKMM}
    S.~Ambrosanio, G.~L.~Kane, G.~D.~Kribs, S.~P.~Martin,
    S.~Mrenna, \PRL{76}{1996}{3498}.

\bibitem{sugra}
     E.~Cremmer, S.~Ferrara, L.~Girardello, A.~van~Proyen,
     \NPB{212}{1983}{413}.

\bibitem{HaberKane}
    H.~E.~Haber, G.~L.~Kane, \PREP{117}{85}{75};
    J.~F.~Gunion, H.~E.~Haber, \NPB{272}{1986}{1};
    Erratum ibid {\bf B402} (1993) 567.

\bibitem{Equivalence}
    J.~M.~Cornwall, D.~N.~Levin, G.~Tiktopoulos, \PRD{10}{1974}{1145},
    Erratum ibid {\bf D11} (1975) {972}.

\bibitem{DicusPRD41}
    D.~A.~Dicus, S.~Nandi, J.~Woodside, \PRD{41}{1990}{2347}.

\bibitem{DreesWoodside}
    M.~Drees, J.~Woodside in Supersymmetry, Proc. of ECFA Workshop
    on the Large Hadron Collider, 1990.

\bibitem{DEJS}
    M.~Drees, J.~Ellis, P.~Jetzer, D.~W.~Sciama,
    \PLB{220}{1989}{586}

\bibitem{spythia} H.U.~Bengtsson, T.~Sj\"{o}strand, Comp. Phys.
Comm. {\bf 46} (1987) 43;
T.~Sj\"{o}strand, Comp. Phys. Comm. {\bf 82} (1994) 74;
S.~Mrenna, ``Simulating Supersymmetry With
{\tt PYTHIA 5.7} and {\tt JETSET 7.4},'' CITHE--68--1987,
July  1995.

\bibitem{baer} See for example
H.~Baer, C.-h.~Chen, C.~Kao, X.~Tata, Phys.~Rev.~{\bf
D52} (1995) 1565; S.~Mrenna, G.L.~Kane, G.D.~Kribs, J.D.~Wells,
Phys.~Rev.~{\bf D53} (1996) 1168.

\bibitem{bob_blair} We thank Bob Blair (Argonne) for providing
information on the CDF diphoton cross-section measurements.

\bibitem{torbjorn} We thank T.~Sj\"{o}strand (Lund) for this
suggestion.

\bibitem{sleptonstev}
    H.~Baer, C.-h.~Chen, F.~Paige, X.~Tata, \PRD{49}{1994}{3283}.

\bibitem{RB} A.~Djouadi et al. \NPB{349}{1991}{48};
M.~Boulware, D.~Finnell, \PRD{44}{1991}{2054};
J.~D.~Wells, C.~Kolda, G.~L.~Kane, \PLB{338}{1994}{219};
D.~Garcia, J.~Sola, \PLB{354}{1995}{335};
G.~L.~Kane, R.~G.~Stuart, J.~D.~Wells, \PLB{354}{1995}{350};
A.~Dabelstein, W.~Hollik, W.~M\"osle, hep-ph/9506251;
P.~H.~Chankowski, S.~Pokorski, \PLB{366}{1996}{188};
J.~Ellis, J.~Lopez, D.~Nanopoulos, \PLB{372}{1996}{95};
J.~D.~Wells, G.~L.~Kane, \PRL{76}{1996}{869};
E.~H.~Simmons, Y.~Su, hep-ph/9602267;
P.~H.~Chankowski, S.~Pokorski, hep-ph/9603310.

%
\bibitem{claes} D.R.~Claes, $10^{\rm th}$ Topical
    Workshop on Proton--Anti-proton Collider Physics, edited by
    Rajendran Raja and John Yoh, AIP Press, 1995,
FERMILAB--CONF--95/186--E.
%
\bibitem{MartinVaughn}
    S.~P.~Martin, M.~T.~Vaughn, \PLB{318}{1993}{331}. The corresponding
(smaller) corrections for charginos and neutralinos are given in
D.~Pierce and A.~Papadopoulos, \PRD{50}{1994}{565},
\NPB{430}{1994}{278}.

\bibitem{BCKT}
    H.~Baer, C.-h.~Chen, C.~Kao, and X.~Tata, \PRD{52}{1995}{1565}.

\bibitem{detectorcuts}
F.~Boudjema, B.~Mele (Conveners),
E.~Accomando et al.~[The Standard Model Process Group],
Proc. of the Workshop ``Physics at LEP2",
G.~Altarelli, T.~Sj\"ostrand, F.~Zwirner (eds.),
CERN-Report 96-01 (1996),
hep-ph/9601224.

\bibitem{CompHEP}
    E.~E.~Boos, M.~N.~Dubinin, V.~A.~Ilin, A.~E.~Pukhov, V.~I.~Savrin,
    ``CompHEP: Specialized package for automatic calculation of
    elementary particle decays and collisions'', hep-ph/9503280, and
    references therein.

\bibitem{DGP}
     G.~Dvali, G.~F.~Giudice, A.~Pomarol, hep-ph/9603238.

\bibitem{BMY}
     S.~Borgani, A.~Masiero, M.~Yamaguchi, hep-ph/9605222.
It is not clear that the requirements of this paper can be satisfied
if the CDF $ee\gamma\gamma\Et$ event is interpreted as superpartner
production.

\end{thebibliography}
\end{document}